\documentclass[12pt]{article}
\pdfoutput=1

\usepackage{pgfplots}
\usetikzlibrary{decorations.pathmorphing}

\tikzset{snake it/.style={decorate, decoration=snake}}
\pgfplotsset{compat=1.10}
\usepgfplotslibrary{fillbetween}
\usetikzlibrary{patterns}

\usepackage{draft} 
\usepackage{mathabx}
\usepackage{hyperref}
\usepackage{graphicx,color,subfig}
\usepackage{cite}
\usepackage{mciteplus}
\usepackage{skak}
\usepackage{xcolor}
\usepackage{empheq}
\usepackage{tikz}
\DeclareFontFamily{OT1}{pzc}{}
\DeclareFontShape{OT1}{pzc}{m}{it}{<-> s * [1.10] pzcmi7t}{}
\DeclareMathAlphabet{\mathpzc}{OT1}{pzc}{m}{it}

\def\be#1\ee{\begin{align}#1\end{align}}



\newcommand{\IE}[0]{{\textit{i.e. }}}

\begin{document}

\unitlength = .8mm

\begin{titlepage}

\begin{center}

\hfill \\
\hfill \\
\vskip 1cm

\title{\Huge Geometry of Conformal Manifolds and the Inversion Formula}

\author{Bruno Balthazar and Clay C\'ordova}
\address{
Kadanoff Center for Theoretical Physics and Enrico Fermi Institute\\ University of Chicago, Chicago IL 60637
}
\vskip 1cm

\email{brunobalthazar@uchicago.edu, clayc@uchicago.edu}

\end{center}

\abstract{Families of conformal field theories are naturally endowed with a Riemannian geometry which is locally encoded by correlation functions of exactly marginal operators.   We show that the curvature of such conformal manifolds can be computed using Euclidean and Lorentzian inversion formulae, which combine the operator content of the conformal field theory into an analytic function. Analogously, operators of fixed dimension define bundles over the conformal manifold whose curvatures can also be computed using inversion formulae. These results relate curvatures to integrated four-point correlation functions which are sensitive only to the behavior of the theory at separated points.  We apply these inversion formulae to derive convergent sum rules expressing the curvature in terms of the spectrum of local operators and their three-point function coefficients.  We further show that the curvature can smoothly diverge only if a conserved current appears in the spectrum, or if the theory develops a continuum.  We verify our results explicitly in $2d$ examples. In particular, for $2d$ (2,2) superconformal field theories we derive a lower bound on the scalar curvature, which is saturated by free theories when the central charge is a multiple of three.

}

\vfill

\begin{flushleft}
December, 2022
\end{flushleft}

\end{titlepage}

\eject

\begingroup
\hypersetup{linkcolor=black}
\setcounter{tocdepth}{2}
\tableofcontents
\endgroup

\section{Introduction and Summary}

The dynamics of many different physical systems are captured by conformal field theories (CFTs). Examples include second order phase transitions of matter, the fixed points of renormalization group flows of quantum field theories, and theories of quantum gravity. In many of these applications the CFT is not isolated, but instead naturally occurs in a family, the \emph{conformal manifold}, related by tuning coupling constants.  As these parameters vary, the data of the CFT, such as its scaling dimensions and coupling constants change continuously thus giving a multi-parameter solutions to the crossing equations. In this paper we explore the local geometry of such families and clarify how the intrinsic data of the CFT is related to natural geometric quantities on the space of theories such as the Riemann curvature.  

Our main results are expressions for the curvature of various bundles over the conformal manifold in terms of the inversion formula for CFT 4-point functions \cite{Caron-Huot:2017vep}.  These formulas are universal: they depend only on correlation functions at separated points.  We use these inversion formulae to obtain convergent sum rules relating the curvature of the conformal manifold to CFT data.  This clarifies and unifies previous work of \cite{Kutasov:1988xb,Ranganathan:1992nb,Ranganathan:1993vj,deBoer:2008ss,Friedan:2012hi}, where expressions for the curvature as integrated four-point correlation functions were also obtained.  We also illustrate our results in several examples in two-dimensions.

\subsection{Families of Conformal Field Theories}

Let us review the basic features of conformal manifolds \cite{Kutasov:1988xb}.  The essential point is that local data of the conformal manifold is encoded by the CFT operators and their correlation functions.  The family is controlled by exactly marginal operators ${\cal O}_i(x)$, $i=1,\cdots N$. Deforming the initial CFT by exponentiating the exactly marginal operators yields a new CFT with different correlation functions.  Thus, the correlation functions in the deformed CFT are formally given by
\begin{equation}
    \begin{split}
        \left\langle \varphi_{I_1}(x_1)\cdots \varphi_{I_n}(x_n)\right\rangle_{\lambda}=Z_\lambda^{-1}\langle \varphi_{I_1}(x_1)\cdots \varphi_{I_n}(x_n)\exp\left(\int \frac{d^d x}{S_{d-1}} \lambda^i {\cal O}_i(x)\right)\rangle~,
    \end{split}
    \label{conf_mfld_corr_CPT1}
\end{equation}
where $\varphi_{I}$ are generic operators in the undeformed CFT, $S_{d-1}$ is the volume of the $(d-1)$-sphere, $\lambda^i$ are couplings, and here and below correlation functions with the $\lambda$ subscript omitted refer to those at the reference CFT at vanishing $\lambda$. In \eqref{conf_mfld_corr_CPT1} the vacuum correlation function $Z_\lambda$ is formally defined by
\begin{equation}
    \begin{split}
        Z_\lambda=\langle \exp\left(\int \frac{d^d x}{S_{d-1}} \lambda^i {\cal O}_i(x)\right)\rangle~.
    \end{split}
\end{equation}
The space of CFTs continuously connected in this way is called the conformal manifold. The couplings $\lambda^i$ give a coordinate chart on the conformal manifold of CFTs, and the operators ${\cal O}_i(x)$ can be thought of as tangent vectors in this manifold \cite{Kutasov:1988xb}.  Correlation functions can be computed order by order in $\lambda$ by expanding the exponential in \eqref{conf_mfld_corr_CPT1} using conformal perturbation theory. A regularization scheme is necessary due to divergences in the integrals when operators collide. This regularization will be discussed in more detail below.

To preserve conformal invariance at leading order in $\lambda$, the operators ${\cal O}_i(x)$ must have dimension $\Delta$ equal to the spacetime dimension $d$, and vanishing spin $J$. At higher order in $\lambda$, we require that the exactly marginal operators in the deformed theory \eqref{conf_mfld_corr_CPT1} also preserve conformal invariance, \IE
\begin{equation}
    \begin{split}
        \frac{\partial}{\partial \ln |x|} |x|^{2d}\left\langle {\cal O}_i(x){\cal O}_j(0)\right\rangle_\lambda=0~.
    \end{split}
    \label{2pt_def}
\end{equation}
For instance, at next-to-leading order in $\lambda$, \eqref{2pt_def} implies that the three-point function coefficeint between exactly marginal operators, $C_{ijk},$ vanishes.

As defined above, the conformal manifold is locally smooth and Riemannian (though these features can both breakdown at special loci). In particular it is endowed with a natural Riemannian metric introduced by Zamolodchikov \cite{Zamolodchikov:1986gt}:
\begin{equation}
    \begin{split}
        g_{ij}(\lambda)\equiv\left\langle {\cal O}_i(e){\cal O}_j(0)\right\rangle_\lambda~,
    \end{split}
    \label{zammet}
\end{equation}
where $e=(1,0,\cdots,0)$. The Zamolodchikov metric $g_{ij}$ is positive definite by unitarity, which we assume henceforth. To compare correlation functions in nearby CFTs using \eqref{conf_mfld_corr_CPT1}, we have to specify a connection on the conformal manifold. In the CFT, this is encoded in a contact term between the exactly marginal operators \cite{Seiberg:1988pf,Kutasov:1988xb,Ranganathan:1992nb,Ranganathan:1993vj}, 
\begin{equation}
    \begin{split}
        {\cal O}_i(x){\cal O}_j(0)\supset\Gamma^k_{ij}(\lambda) {\cal O}_k(0)\delta^{(d)}(x)~,
    \end{split}
    \label{christ}
\end{equation}
where $\Gamma^k_{ij}(\lambda)$ are the Christoffel symbols of the metric \eqref{zammet}. At leading order in $\lambda$, the contact terms are not specified by the CFT data, which governs correlation functions at separated points. Instead, they are specified by a choice of counterterms to the action \eqref{conf_mfld_corr_CPT1}, which are needed once we deform the theory.

The choice of counterterms is intimately tied to the choice of local coordinates $\lambda^i$. More precisely, changing coordinates $\lambda^i\to \tilde\lambda^i$ is equivalent to a change in the contact term \eqref{christ}, in such a way that \eqref{christ} changes to the Christofell symbol in the coordinate system $\tilde\lambda$. In particular, we can always choose to work in local coordinates where the contact terms vanish locally, $\Gamma_{ij}^k(0)=0$.

While we can choose contact terms to vanish locally, their derivative is not necessarily zero, and captures the Riemannian curvature of the conformal manifold. In coordinates where $\Gamma^{k}_{ij}(0)=0$, this is given by the simple expression:
\begin{equation}
    \begin{split}
R^i{}_{jk\ell}=\partial_k\Gamma_{\ell j}^i-\partial_\ell\Gamma_{kj}^i~.
    \end{split}
    \label{Rintro}
\end{equation}
One of our main aims below is to derive an expression for this curvature in terms of CFT correlation functions, and to explain how such expressions relate the curvature to spectrum of operators and their OPE coefficients.  

The ideas above can also be extended to define the geometry of other bundles on the conformal manifold.  Consider the operators $\varphi_I(x)$, $I=1,\cdots,L$, of dimension $\Delta$ and spin $J$ and further assume that as we vary $\lambda$ these quantum numbers stay constant. (This is a natural assumption for instance if the operators are currents or BPS.) Under deformation by ${\cal O}_i(x)$, the operators $\varphi_I(x)$ can mix with each other, \IE they define sections of vector bundles over the conformal manifold \cite{deBoer:2008ss,Papadodimas:2009eu}. The contact terms between $\varphi_I(x)$ and ${\cal O}_i(x)$ give connections for these bundles,
\begin{equation}
    \begin{split}
        {\cal O}_i(x){\varphi}_J(0)\supset A_i{}^{K}{}_{J} (\lambda){\varphi}_K(0)\delta^{(d)}(x)~.
    \end{split}
    \label{Acon}
\end{equation}
These contact terms can also be adjusted by tuning counterterms in the action. While we can set them to zero at a given point in the conformal manifold, their derivative is in general non-vanishing. Hence, these bundles also have curvature, which is given by
\begin{equation}
    \begin{split}
        F_{ij}{}^{K}{}_L\equiv\partial_{i}A_j{}^{K}{}_{L} -\partial_{j}A_i{}^{K}{}_{L}~. 
    \end{split}
    \label{Fintro}
\end{equation}
We will see that this curvature also admits an expression in terms of local CFT data.

\subsection{Examples of Conformal Manifolds and Motivation}

We will now review some general features of conformal manifolds in the literature. All known examples of CFTs with exactly marginal operators have extra symmetries. These symmetries ensure the exact marginality of the operators ${\cal O}_i(x)$ at any point in the conformal manifold. Below, we will not make any assumptions about the existence of extra symmetries, unless explicitly stated.
 
 Free theories in two-dimensions provide the simplest examples of conformal manifolds. It is known from the Narain construction of these theories that the conformal manifold is a homogeneous space with constant non-positive  Riemann curvature (see \cite{Polchinski:1998rq} and references therein). There can be orbifold singularities at points with enhanced symmetry, such as self-dual points, as well as cusp singularities, such as the infinite volume limit. We will always assume that we are working at a generic point in the conformal manifold where the local geometry is smooth.

Apart from theories in two-dimensions with extended chiral symmetry, the only other known examples of conformal manifolds arise in supersymmetric theories. In two-dimensions, this includes $(2,2)$ SCFTs such as non-linear sigma models on Calabi-Yau manifolds. Exactly marginal deformations are in 1-to-1 correspondence with the Hodge numbers of the Calabi-Yau space \cite{Witten:1981nf}. Such conformal manifolds can also exhibit other types of singularities, such as conifold singularities \cite{Candelas:1989ug,Candelas:1989js}. Near these singularities, the Riemann curvature becomes positive and diverges \cite{Candelas:1990rm}.

There are also myriad examples of conformal manifolds in supersymmetric theories in higher dimensions. This includes $4d$ ${\cal N}=4$ super Yang-Mills, where the complexified gauge coupling parametrizes the conformal manifold. In this case, the weak coupling limit $\tau\to i\infty$ is a cusp-like singularity. A zoo of other examples with $4d$ ${\cal N}=2$ or ${\cal N}=1$ or $3d$ $\mathcal{N}=2$ supersymmetry are also known (see e.g.\ \cite{Gaiotto:2009we}). In this context the local structure of the conformal manifold can be studied using superconformal representation theory \cite{Leigh:1995ep, Green:2010da}.  With sufficient supersymmetry, the Zamolodchikov metric can be computed via supersymmetric localization techniques \cite{Jockers:2012dk, Gomis:2012wy, Gerchkovitz:2014gta, Gomis:2014woa, Gomis:2015yaa}.   It is also known that all supersymmetric examples of conformal manifolds necessarily occur in spacetime dimension $d\leq 4$ as supersymmetric CFTs with $d=5$ and $d=6$ do not admit exactly marginal deformations  preserving the supercharges.  Similarly theories with $d=3$ and $\mathcal{N}>2$ do not admit supersymmetric exactly marginal operators \cite{Cordova:2016xhm, Cordova:2016emh}. 

One motivation of this work is to derive properties of CFTs using the curvature of the conformal manifold as input. In the examples above, the curvature of the conformal manifold is computed from alternative methods that do not rely on the correlation functions of the CFT. The question we want to ask is the following: given some information about the geometric properties of a conformal manifold of a CFT, what are the implications for the CFT data? This setup gives new constraints on the CFT data, which can be used in combination with other techniques such as the conformal bootstrap (see \cite{Hartman:2022zik,Poland:2022qrs} and references therein) to further constrain these theories. Relatedly, the integrated correlation functions characterizing the curvature may be utilized in the conformal bootstrap-like analysis as in \cite{Lin:2015wcg,Lin:2016gcl,Chester:2020vyz, Chester:2021aun,Collier:2022emf}.

There have also been conjectures relating the geometry of conformal manifolds to the operator spectrum of CFTs in $d>2$ dimensions \cite{Perlmutter:2020buo}. In particular, it is conjectured that at infinite distance in the conformal manifold (with respect to the Zamolodchikov metric \eqref{zammet}), a tower of higher spin operators with unbounded spin saturates the unitarity bound, thus becoming conserved currents. This is indeed observed in all examples of (super)conformal manifolds. Another motivation of this work is to build general tools in CFT to potentially address such relationships between conformal manifolds and aspects of the CFT spectrum.

\subsection{The Inversion Formula and Results}

Given a four-point function in a CFT, the inversion formula outputs a function $I_{\Delta,J}$ of dimension $\Delta$ and spin $J$ \cite{Caron-Huot:2017vep,Simmons-Duffin:2017nub,Kravchuk:2018htv}. For integer spin $J$, the function $I_{\Delta,J}$ has poles in $\Delta$ corresponding to operators in the $s$-channel OPE expansion. The location of the pole gives the dimension of the exchanged operator, while the residue encodes the structure constants.

There are two types of inversion formulae, a Euclidean inversion formula and a Lorentzian inversion formula. The input of the Lorentzian inversion formula is a double-discontinuity of the correlation function, and unlike the Euclidean inversion formula, it outputs a function that is analytic in spin. It has been used to obtain large spin corrections to the anomalous dimensions of operators, with controlled errors \cite{Caron-Huot:2017vep}. It has also been used to prove the average null energy condition \cite{Kravchuk:2018htv}.

The main result of this work is a relation between the curvature of vector bundles over the conformal manifold \eqref{Fintro} and the inversion formula $I_{\Delta,J}$. In section \ref{sec:inversion}, we derive our main result\footnote{Note that, since $I_{\Delta,J}=I_{d-\Delta,J}$ \cite{Simmons-Duffin:2017nub}, the curvature is finite if there are no conserved current exchanged in the $s$-channel. The presence of such currents would imply that the operators ${\cal O}_i$ are charged under a global symmetry which is broken by the deformation. It has been shown that charged marginal operators are not exactly marginal, see \cite{Friedan:2012hi} for the discussion in $2d$ CFTs, and \cite{Green:2010da} for $d>2$ SCFTs.}
\begin{equation}
    \begin{split}
        F_{ij}{}^K{}_L&= \frac{{\text{vol}}(SO(d-1))}{S_{d-1}^2} I_{\Delta=1,J=1}~,
    \end{split}
    \label{Finv_intro}
\end{equation}
where the Euclidean inversion formula appearing in this formula is given by
\begin{equation}
    \begin{split}
        I_{\Delta=1,J=1}=\frac{1}{2}\int \frac{d^2z}{{\text{vol}}(SO(d-1))} \left|\frac{z-\bar z}{2i}\right|^{d-2}  \ln|1-z|^2\langle \varphi^{K}(0)\varphi_{L}(z,\bar z){\cal O}_{i}(1){\cal O}_{j}(\infty)\rangle_c~,
    \end{split}
    \label{Iintro}
\end{equation}
and above we assume $\varphi_I$ to be scalar operators for simplicity. Here, indices are raised and lowered using the 2-point function of the operators, $\left\langle\varphi_I(1)\varphi_J(0)\right\rangle$. Furthermore, the notation $\langle\cdots\rangle_c$ denotes the connected part of the correlation function, \IE with the vacuum contribution in the $s,t,u$-channels stripped off.

The $2d$ Euclidean inversion formula had been obtained previously in the literature using the Weyl anomaly \cite{Friedan:2012hi}. The expression \eqref{Iintro} generalizes this result to higher dimensions, and allows us to write the curvature as an integral over a Lorentzian domain by using the Lorentzian inversion formula.

Using the Lorentzian inversion formula, we can derive a sum rule for the curvature $F_{ij}{}^K{}_L$. For the curvature of the tangent bundle in $2d$, \IE the Riemann curvature of the Zamolodchikov metric, this is given by
\begin{equation}
    \begin{split}
        R_{ijk\ell}&=\frac{1}{8\pi^2}\sum_{A}\left(C_{\ell i}{}^AC_{kj A}-C_{\ell j}{}^AC_{ki A}\right)\frac{\left[1-\cos\left(2\pi \bar{h}_A\right)\right]}{(1-h_A)^2(1-\bar h_A)^2}~.
    \end{split}
    \label{Fsumrule_intro}
\end{equation}
The sum is over all operators $V_A$ with conformal weights $(h_A,\bar h_A)$, and $C_{ij}{}^A$ is the structure constant $\left\langle{\cal O}_i(0){\cal O}_j(e) V^A(\infty)\right\rangle$. In $4d$, the sum rule for the Riemannian curvature is given by
\begin{equation}
    \begin{split}
       R_{ijk\ell}&=\frac{32}{\pi^2}\sum_{A}\left(C_{jk A}C_{i\ell A}-C_{j\ell A}C_{ik A}\right)\left[1-\cos\left(\pi(\Delta_A+J_A)\right)\right]
        \\
        &\times\frac{7J_A^4-2(44+3(\Delta_A-8)\Delta_A)J_A^2-(\Delta_A-6)^2(\Delta_A-2)^2}{\left[(\Delta_A-2)^2-J_A^2\right]^2\left[(\Delta_A-4)^2-J_A^2\right]^2\left[(\Delta_A-6)^2-J_A^2\right]^2}~.
    \end{split}
    \label{Fsumrule_intro_4d}
\end{equation}
where $\Delta_A$ and $J_A$ are the dimension and spin of the operator $V_A$.\footnote{It is possible to rewrite these sums over global conformal primaries only, but the result does not look simpler.} The sum over operators $V_A$ converges provided that there are no higher spin conserved currents in the $s$-channel OPE. There are similar sum rules in other dimensions.  The contributions from operators of high dimensions and fixed spin are suppressed in the sum, suggesting that such formulas may be useful tools for constraining the properties of light operators.  

The inversion formula is invariant under the shadow transformation $\Delta\to \tilde\Delta\equiv d-\Delta$, which maps the quantum numbers $\Delta=1,J=1$ to $\tilde\Delta=d-1,J-1$. These are the quantum numbers of a conserved current. Using this together with \eqref{Finv_intro}, we conclude that if the curvature $F_{ij}{}^K{}_L$ diverges at a certain point in moduli space, it must be that either a conserved current appears in the $s$-channel OPE, or the analyticity of the inversion formula $I_{\Delta,J}$ breaks down. This can happen if the spectrum of operators in the CFT develops a continuum. Note that in $d=2$ the only way the curvature can diverge is if the CFT develops a continuum, since a pole at $\Delta=1$ is mapped to minus itself under the shadow transformation.\footnote{The divergence of the curvature we are considering here does not include orbifold singularities in the conformal manifold, which are delta-function localized.}

We will further derive an inversion formula for the second derivative of the Zamolodchikov metric, $g_{ij,k\ell}$. Since this quantity does not transform as a tensor under a coordinate change, it is only meaningful once we specify our choice of counterterms to the action \eqref{conf_mfld_corr_CPT1}. We will show that in the minimal subtraction scheme, where we set the finite part of these counterterms to zero, we are working in Riemann normal coordinates, and $g_{ij,k\ell}$ is related to the inversion formula $I_{\Delta=0,J=0}$.

Using this result, we derive new sum rules for the Riemannian curvature of the conformal manifold. In this case, convergence of the sum rules requires the absence of more operators in the OPE of exactly marginal operators, not just higher spin conserved currents. Assuming these operators to be absent in the OPE, we derive some simple bounds in the sectional curvature of the conformal manifold. For example, if all the operators exchanged by exactly marginal operators in $4d$ have dimension and spin in the range \eqref{4dbadops}, then the sectional curvature of the conformal manifold satisfies
\begin{equation}
    \begin{split}
        R_{ijij}\geq0~.
    \end{split}
    \label{bound4dsimple_intro}
\end{equation}
Note that if there are operators outside the range \eqref{4dbadops} that are present in the OPE of exactly marginal operators, their contribution can be explicitly included in the sum rules. This will in general modify the RHS of \eqref{bound4dsimple_intro}.  Conversely, in many examples, such as in neighborhoods of weakly-coupled gauge theories characterized by cusps, the sectional curvature is known to be negative and hence we can conclude that operators outside the range \eqref{4dbadops} necessarily exist.

Finally, in section \ref{sec:examples}, we exhibit the master formulae \eqref{Finv_intro},\eqref{Iintro} in a number of examples in two-dimensions where the four-point function of exactly marginal operators is known explicitly. First we will consider free theories, where we will recover the metric on a homogeneous space of negative constant curvature. Then we will consider $2d$ $(2,2)$ SCFTs, where we will use our methods to show that the curvature of the vector bundles of chiral primary operators over the conformal manifold agrees with that obtained from the $t t^*$ equations \cite{Cecotti:1991me,Bershadsky:1993cx}. In particular, we find that the scalar curvature of $2d$ (2,2) SCFTs satisfies the lower bound
\begin{equation}
\begin{split}
        R\geq -\frac{n(n+1)}{4}~,
\end{split}
\end{equation}
where $n$ is the (complex) dimension of the conformal manifold. When $c/3$ is an integer, this bound is saturated by the CFT of $c/3$ complex bosons+fermions.

\section{Conformal Manifolds in Perturbation Theory}
\label{sec:conf_mfld_rev}

In this section, we review properties of conformal manifolds in more detail. We discuss how correlation functions change as we deform the CFT, organizing the discussion order by order in the conformal perturbation theory expansion of \eqref{conf_mfld_corr_CPT1}.

\subsection{Conformal Perturbation Theory at Leading Order}

Let's begin by reviewing the result that the three-point function of exactly marginal operators vanishes, discussed below \eqref{2pt_def}. Along the way, we will also define the regularization procedure we use to define \eqref{conf_mfld_corr_CPT1}.

 Under the deformation \eqref{conf_mfld_corr_CPT1}, the two-point function of exactly marginal operators is corrected at leading order in $\lambda$ by
\begin{equation}
         \partial_k\left\langle{\cal O}_i(x_1){\cal O}_j(x_2)\right\rangle_\lambda|_{\lambda=0}=\int \frac{d^d x_3}{S_{d-1}}\left\langle{\cal O}_i(x_1){\cal O}_j(x_2){\cal O}_k(x_3)\right\rangle
        =\frac{C_{ijk}}{(x^2_{12})^\frac{d}{2}}\int \frac{d^d x_3}{S_{d-1}} \frac{1}{(x^2_{13})^\frac{d}{2}(x^2_{23})^\frac{d}{2}}~,
    \label{leading_anom_dim}
\end{equation}
where $\partial_k\equiv{\partial}/{\partial\lambda^k}$, $C_{ijk}\equiv \langle {\cal O}_i(\infty){\cal O}_j(1){\cal O}_k(0)\rangle$, and we are dropping terms of order ${\cal O}(\lambda)$ on the RHS. The integral on the RHS of \eqref{leading_anom_dim} is divergent at coincident points where $x^2_{13}=0,x^2_{23}=0$. This is a recurring theme in conformal perturbation theory. The definition of the deformed theory via \eqref{conf_mfld_corr_CPT1} is formally infinite, and needs to be regularized due to divergences in the integrals when operators collide. 

There are two common regularization schemes. One is hard-sphere regularization, in which spherical neighbourhoods of radius $\epsilon$ around the singular points $x^2_{13}=0,x^2_{23}=0$ are cut out. The integrals are performed with this UV cutoff, and in the end we adjust counterterms to cancel the divergences in $\epsilon$.

Another convenient choice of regularization scheme, which is the one we will use, is analytic regularization, utilized in a similar context in \cite{Friedan:2012hi}.  In this case, instead of the correlation function of operators $\left\langle \varphi_{I_1}(x_1)\cdots \varphi_{I_n}(x_n)\right\rangle$, we consider the regulated correlation function:
\begin{equation}
    \begin{split}
        G^{}_{s}(x_1,\cdots,x_n)\equiv\mu^{n s}\left(\prod_{\alpha<\beta}(x^2_{\alpha\beta})^{\frac{s}{n-1}}\right)\left\langle \varphi_{I_1}(x_1)\cdots \varphi_{I_n}(x_n)\right\rangle~,
    \end{split}
    \label{reg_npt}
\end{equation}
where $\mu$ is a mass-scale, and for simplicity we dropped the indices on the LHS. Integrals are performed for sufficiently large $s$, and then analytically continued to $s=0$. Any poles in $s$ are cancelled by properly adjusting the counterterms. Note that, logarithmic contributions in $\mu$ appear if the integral gives a pole in $s$. These logarithmic contributions are due to the anomalous dimension of the operators.

Using analytic regularization in \eqref{leading_anom_dim}, we find
\begin{equation}
    \begin{split}
 \partial_k\left\langle{\cal O}_i(x_1){\cal O}_j(x_2)\right\rangle_\lambda|_{\lambda=0}= \mu^{3s}\frac{C_{ijk}}{(x^2_{12})^{\frac{d-s}{2}}}\int \frac{d^d x_3}{S_{d-1}} \frac{1}{(x^2_{13})^{\frac{d-s}{2}}(x^2_{23})^{\frac{d-s}{2}}}~.
    \end{split}
    \label{2pt_CPT_full}
\end{equation}
The logarithmic dependence on the scale $\mu$ comes from a pole in $s$ in the integral, which can come from the region $x_3$ close to $x_1$ or $x_2$. Performing the integral in these regions, we find
\begin{equation}
    \begin{split}
        \partial_k\left\langle{\cal O}_i(x_1){\cal O}_j(x_2)\right\rangle_\lambda|_{\lambda=0}=6 \frac{C_{ijk}}{(x^2_{12})^d} \ln\left(\mu|x_{12}|\right)+\text{(analytic in $\mu$ as $s\to0$)}~.
    \end{split}
    \label{2pt_CPT_correction}
\end{equation}
Thus, up to terms analytic in $\mu$, we find that the two-point function of exactly marginal operators in the deformed theory is given by
\begin{equation}
    \begin{split}
        \left\langle{\cal O}_i(x_1){\cal O}_j(x_2)\right\rangle_{\lambda}=\frac{g_{ij}(0)}{(x_{12}^2)^d}+6\lambda^k \frac{C_{ijk}}{(x^2_{12})^d} \ln\left(\mu|x_{12}|\right)+{\cal O}(\lambda^2)~.
    \end{split}
    \label{2pt_CPT_all}
\end{equation}
Comparing this result to a two-point function of a generic operator of dimension $\Delta$, we see that the dimension of ${\cal O}_i$ is modified at leading order to
\begin{equation}
    \begin{split}
        \Delta=d-3\lambda^k \frac{C_{ijk}}{g_{ij}(0)}+{\cal O}(\lambda^2)~.
    \end{split}
    \label{Oi_anomdim}
\end{equation}
Hence, we find the well known result, that exact marginality implies $C_{ijk}=0$.

After regularization, we also have to specify counterterms to the deformed CFT action. A convenient way of encoding this information is as follows. We promote the couplings $\lambda^i$ to background fields $\lambda^i(x)$. These fields source the exactly marginal operators ${\cal O}_i$, so we can define insertions of ${\cal O}_i$ via functional differentiation with respect to $\lambda^i(x)$, \IE
\begin{equation}
    \begin{split}
       \langle {\cal O}_i(x) \cdots\rangle_{\lambda}\to \frac{\delta}{\delta \lambda^i(x)}\langle  \cdots\rangle_{\lambda}~,
    \end{split}
\end{equation}
where $\cdots$ denote operators in the reference CFT at separated positions and away from $x=0$. After evaluating the functional derivatives, we set the background fields $\lambda^i$ to constants. After promoting the couplings $\lambda^i$ to background fields $\lambda^i(x)$, the allowed counterterms are given by contributions to the action that are local in $\lambda^i(x)$. 

Let us show for example that counterterms do not affect the anomalous dimension in \eqref{Oi_anomdim}. This is because the dependence on the scale $\mu$ in \eqref{2pt_CPT_correction} is non-analytic, and cannot be cancelled by local counterterms. At the order of $\lambda$ that we are interested in, we consider the counterterms
\begin{equation}
    \begin{split}
        \int \frac{d^d x}{S_{d-1}} \frac{1}{2}B^{k}_{ij}\lambda^i(x)\lambda^j(x) {\cal O}_k(x)~.
    \end{split}
    \label{ctterms_1}
\end{equation}
In other words, we now compute correlation functions of ${\cal O}_i$, at leading order in $\lambda$, in the theory defined by
 \begin{equation}
    \begin{split}
        &\left\langle {\cal O}_{i_1}(x_1)\cdots {\cal O}_{i_n}(x_n)\right\rangle_\lambda
        \\
        &= Z_\lambda^{-1}\left(S_{d-1}\right)^{n}\frac{\delta}{\delta \lambda^{i_1}(x_1)}\cdots\frac{\delta}{\delta \lambda^{i_n}(x_n)}\langle\exp\left(\int \frac{d^d x}{S_{d-1}}\left( \lambda^i(x) {\cal O}_i(x)+\frac{1}{2}B^{k}_{ij}\lambda^i(x)\lambda^j(x) {\cal O}_k(x)\right)\right)\rangle~,
    \end{split}
    \label{conf_mfld_corr_CPT_ctterms}
\end{equation}
with the regularization described above, and setting $\lambda^i$ to a constant after taking the functional derivatives. The choice of constants $B^k_{ij}$ specifies the counterterms, which are also included in $Z_\lambda$.

With these counterterms, the RHS of \eqref{2pt_CPT_all} gets a contribution equal to
\begin{equation}
    \begin{split}
 \left\langle{\cal O}_i(x_1){\cal O}_j(x_2)\right\rangle_{\lambda}\supset\lambda^k  \frac{1}{(x^2_{12})^d} \left(B^\ell_{ik}g_{\ell j}(0)+B^\ell_{jk}g_{\ell i}(0)\right)~.
    \end{split}
    \label{3pt_ctterms}
\end{equation}
This contribution cannot cancel the non-analytic contribution on the RHS of \eqref{2pt_CPT_all}, but it can cancel any term with analytic dependence on the scale $\mu$ (including the $s$-pole in the integral in \eqref{2pt_CPT_full}). Thus, we find that $C_{ijk}=0$ for exact marginality of the operators ${\cal O}_i(x)$, as discussed above.

The counterterms give a finite contribution \eqref{3pt_ctterms} to the deformed 2-point function of exactly marginal operators. This seems to contradict the fact that observables in the theory are unaffected by our choice of counterterms. However, looking more closely at \eqref{3pt_ctterms}, we see that it also suffers from another ambiguity, namely an ambiguity in the choice of couplings $\lambda^i$. Indeed, at leading order in conformal perturbation theory, the field redefinition:
\begin{equation}
    \begin{split}
        \lambda^i(x)\to \tilde\lambda^i(x)= \lambda^i(x)-B^{i}_{jk}\lambda^j(x)\lambda^k(x)~,
    \end{split}
    \label{coord_change_1}
\end{equation}
absorbs the counterterm in \eqref{conf_mfld_corr_CPT_ctterms}. Thus, the $\lambda^i(x)$ coordinates with the counterterms \eqref{ctterms_1} is equivalent to the choice of coordinates $\tilde\lambda^i(x)$, and counterterms set to zero. In other words, the parameters $\lambda^i$ are coordinates on the conformal manifold, and adjusting counterterms correspond to changing these coordinates.

This discussion has a natural interpretation in terms of the connection of the Zamolodchikov metric \eqref{zammet} of the conformal manifold. Consider the derivative of the Zamolodchikov metric with respect to the coordinate $\lambda^k$, which is given by
\begin{equation}
g_{ij,k}(0)\equiv \left.\partial_k\left\langle{\cal O}_i(x_1){\cal O}_j(x_2)\right\rangle_{\lambda}\right|_{\lambda^i=0}
        = B^\ell_{ik}g_{\ell j}(0)+B^\ell_{jk}g_{\ell i}(0)~,
    \label{gij_dk}
\end{equation}
where in the last line we used \eqref{3pt_ctterms}. Suppose now we change our choice of counterterms, $B^{i}_{jk}\to B^{i}_{jk}+\delta B^{i}_{jk}$. We can compare the variation of \eqref{gij_dk} to that expected for the variation of a Riemannian metric under the coordinate change \eqref{coord_change_1}. The latter is given by
\begin{equation}
    \begin{split}
    \frac{\partial}{\partial \tilde\lambda^k}\left(\frac{\partial \lambda^p}{\partial\tilde\lambda^i}\frac{\partial \lambda^q}{\partial\tilde\lambda^j}g_{pq}\right)(0)&= g_{ij,k}(0)+\frac{\partial ^2\lambda^p}{\partial\tilde\lambda^i\partial\tilde\lambda^k}\frac{\partial \lambda^q}{\partial\tilde\lambda^j} g_{pq}(0)+\frac{\partial \lambda^p}{\partial\tilde\lambda^i}\frac{\partial^2 \lambda^q}{\partial\tilde\lambda^j\partial\tilde\lambda^k}g_{pq}(0)
\\
&= g_{ij,k}(0)+ \left(\delta B^\ell_{ik}g_{\ell j}(0)+\delta B^\ell_{jk} g_{\ell i}(0)\right),
    \end{split}
    \label{gij_dk_coordtransf}
\end{equation}
This transformation is exactly as expected from the conformal perturbation theory analysis in \eqref{gij_dk}.

For a Riemannian metric, we have
\begin{equation}
    \begin{split}
        g_{ij,k}(0)=\Gamma^\ell_{ik}g_{\ell j}(0)+\Gamma^\ell_{jk}g_{\ell i}(0)~.
    \end{split}
    \label{gij_dk_CS}
\end{equation}
Comparing \eqref{gij_dk} and \eqref{gij_dk_CS}, we see that $\Gamma^i_{jk}=B^i_{jk}$. Thus, the counterterms are naturally interpreted as connections on the conformal manifold. This also explains why they modify correlation functions: a change of counterterms is equivalent to a different choice of coordinate chart in the conformal manifold, and therefore we compute correlation functions at different points in the conformal manifold.

It is also useful to think of the counterterms as determining the contact terms. If we consider the 3-point function of exactly marginal operators at points $0,1,x$, we find from \eqref{conf_mfld_corr_CPT_ctterms}
\begin{equation}
    \begin{split}
       \langle {\cal O}_i(e){\cal O}_j(0){\cal O}_k(x)\rangle&=\left(S_{d-1}\right)^3\frac{\delta}{\delta \lambda^{i}(e)}\frac{\delta}{\delta \lambda^{j}(0)}\frac{\delta}{\delta \lambda^{k}(x)}\langle\exp\left(\int \frac{d^d x}{S_{d-1}}\left( \lambda^i(x) {\cal O}_i(x)+B^{k}_{ij}\lambda^i(x)\lambda^j(x) {\cal O}_k(x)\right)\right)\rangle
       \\
       &= S_{d-1}\Gamma^{\ell}_{ik}g_{\ell j}(0)\delta^{(d)}(x-e)+S_{d-1}\Gamma^{\ell}_{jk}g_{\ell i}(0)\delta^{(d)}(x)~.
    \end{split}
    \label{3pt_sep}
\end{equation}
Integrating this result over $x$, we recover \eqref{gij_dk}.

\subsection{Second Order Perturbations and the Riemannian Curvature}

We now consider conformal perturbation theory at second order in the $\lambda^i$ expansion.  We expect to obtain quantities related to the second derivative of the Zamolodchikov metric, out of which we can construct the Riemannian curvature of the manifold. 

To begin we examine the conditions for exact marginality at this order.  Requiring that the 3-point function of exactly marginal operators vanishes in the deformed theory, we find
\begin{equation}
    \begin{split}
        \int \frac{d^d x}{S_{d-1}}\langle{\cal O}_i(\infty){\cal O}_j(e){\cal O}_k(0){\cal O}_\ell(x)\rangle=0~.
    \end{split}
    \label{int_4pt_sep}
\end{equation}
Here we have used the conformal group to place the 3 unintegrated vertex operators at separated points at $0,e\equiv(1,0,\cdots,0),\infty$, and furthermore we take the counterterms $B^k_{ij}$ in \eqref{conf_mfld_corr_CPT_ctterms} to vanish. As discussed above, this corresponds geometrically to a choice of coordinates for the conformal manifold, in which $g_{ij,k}=0$. To regularize the integral, we replace the 4-point function of exactly marginal operators by the regulated 4-point function in \eqref{reg_npt}.

We now turn to the the deformed 3-point function, not necessarily at separated points. Setting the counterterms \eqref{ctterms_1} to zero and ignoring further counterterms for now, we have
\begin{equation}
    \begin{split}
        \int \frac{d^d x_2}{S_{d-1}}\langle{\cal O}_i(e){\cal O}_j(0){\cal O}_k(x_1){\cal O}_\ell(x_2)\rangle=S_{d-1}\partial_\ell\left(\Gamma^{m}_{jk}g_{m i}(0)\right)\delta^{(d)}(x_1)+S_{d-1}\partial_\ell\left(\Gamma^{m}_{ik}g_{m j}(0)\right)\delta^{(d)}(x_1-e)~.
    \end{split}
    \label{int_4pt}
\end{equation}
Here, the RHS is fixed by comparing against \eqref{3pt_sep}, and noting that \eqref{int_4pt} is the leading correction to it under the deformation. Note that the LHS is defined using analytic regularization \eqref{reg_npt}.  Let us compare \eqref{int_4pt} to the second derivative of the metric given by: 
\begin{equation}
         g_{ij,k\ell}(0)= \partial_k\partial_\ell\left\langle{\cal O}_i(e){\cal O}_j(0)\right\rangle_\lambda|_{\lambda=0}
         =\int \frac{d^d x_1}{S_{d-1}}\int \frac{d^d x_2}{S_{d-1}}\langle{\cal O}_i(e){\cal O}_j(0){\cal O}_k(x_1){\cal O}_\ell(x_2)\rangle_c~.
    \label{gij_dkdl}
\end{equation}
In the final equality, we have kept only the connected part of the correlation function (see below \eqref{Iintro}). The identity in the $s$-channel is subtracted because of the vacuum contribution coming from $Z_\lambda$ in \eqref{conf_mfld_corr_CPT1}. The vacuum contribution in the $t,u$-channels can be dropped because it vanishes using analytic regularization, as discussed in section \ref{sec:relops}.  Integrating \eqref{int_4pt} over $x_1$, we find
\begin{equation}
    \begin{split}
        g_{ij,k\ell}(0)=\partial_\ell\left(\Gamma^{m}_{ik}g_{m j}(0)\right)+\partial_\ell\left(\Gamma^{m}_{jk}g_{m i}(0)\right)~,
    \end{split}
\end{equation}
which is indeed correct for a Riemannian metric with locally vanishing first derivative. 

We now turn to the Riemannian curvature of the Zamolodchikov metric.  Assuming locally vanishing first derivatives as above, this is given by
\begin{equation}
    \begin{split}
        R_{ijk\ell}=\frac{1}{2}\left(g_{i\ell,jk}+g_{jk,i\ell}-g_{ik,j\ell}-g_{j\ell,ik}\right)~.
    \end{split}
    \label{Rijkl_gijkl}
\end{equation}
Using \eqref{gij_dkdl}, we see that $ R_{ijk\ell}$ is constructed from the integrated 4-point function of exactly marginal operators, and hence depends on the operators appearing in the OPE of the exactly marginal $\mathcal{O}_{i}$'s. Below we will make this observation sharper.

The Riemann curvature transforms as a tensor under a change of coordinates.  As explained in the previous subsection, the change of coordinates
 \begin{equation}
    \begin{split}
        \lambda^i(x)\to \tilde\lambda^i(x)= \lambda^i(x)-B^{i}_{jk}\lambda^j(x)\lambda^k(x)-C^i_{jk\ell}\lambda^j(x)\lambda^k(x)\lambda^\ell(x)~,
    \end{split}
    \label{coord_change_2}
\end{equation}
is equivalent to computing the 4-point function in the theory deformed by
\begin{equation}
    \begin{split}
        \exp\left(\int \frac{d^d x}{S_{d-1}}\left( \lambda^i(x) {\cal O}_i(x)+\frac{1}{2}\Gamma^{k}_{ij}\lambda^i(x)\lambda^j(x) {\cal O}_k(x)+\frac{1}{3!}C^\ell_{ijk}\lambda^i(x)\lambda^j(x)\lambda^k(x){\cal O}_\ell(x)\right)\right).
    \end{split}
    \label{ctterms_all}
\end{equation}
Using this action, one finds that the ${\cal O}(\lambda^3)$ counterterms contribute a constant proportional to $C^{\ell}_{ijk}$ to $g_{ij,k\ell}$ in \eqref{gij_dkdl}, but this contribution vanishes when taking the symmetrized expression \eqref{Rijkl_gijkl}. Meanwhile, the contribution of the ${\cal O}(\lambda^2)$ counterterm in \eqref{ctterms_all} gives a contribution $\sim \Gamma^2$ to \eqref{gij_dkdl}, and in the Riemannian curvature \eqref{Rijkl_gijkl} it reproduces the expected dependence on the Christofell symbols.

\subsection{Higher Orders and Other Operators}

The geometry of the conformal manifold can be explored at higher orders in conformal perturbation theory. From the $n$-point function of exactly marginal operators, we can extract the $(n-2)$-th metric derivative. Derivatives of the metric do not transform covariantly under change of coordinates, and hence they are affected by the choice of counterterms. However, we can combine these expressions to obtain geometric properties of the conformal manifold, such as covariant derivatives of the Riemann curvature. 

So far, we have focused on correlation functions of exactly marginal operators.  Correlation functions of other operators in the reference CFT are also modified as the CFT is deformed along the conformal manifold. This is also captured by conformal perturbation theory. It is convenient to introduce sources for every operator in the CFT, and define operator insertions as derivatives with respect to the sources. Then, there are new counterterms that are allowed, which are given by terms in the action that are local in all sources.

As an example, consider an operator $\varphi_I(x)$ with dimension and spin $(\Delta,J)$. Introduce the source $a^I(x)$, so that
\begin{equation}
    \begin{split}
        \langle \varphi_I(x)\cdots\rangle_{\lambda,a}\to S_{d-1}\frac{\delta}{\delta a^I(x)}\langle \cdots\rangle_{\lambda,a}~,
    \end{split}
\end{equation}
where we take $\lambda^i(x)$ and $a^I(x)$ to a constant and to zero after taking the functional derivative, respectively. Note that $\varphi_I(x)$ is not marginal, and therefore $a^I(x)$ has dimension $d-\Delta$. Let's consider the correction to the two-point function of scalar operators $\varphi_I, \varphi_J$ of the same dimension $\Delta$. Under the deformation \eqref{conf_mfld_corr_CPT1}, the same steps leading to \eqref{2pt_CPT_all} give
\begin{equation}
    \begin{split}
        \left\langle \varphi_I(x_1)\varphi_J(x_2)\right\rangle_{\lambda}=\frac{h_{IJ}(0)}{(x_{12}^2)^\Delta}+6\lambda^k \frac{C_{IJk}}{(x^2_{12})^\Delta} \ln\left(\mu|x_{12}|\right)+{\cal O}(\lambda^2)~,
    \end{split}
    \label{2pt_CPT_allops}
\end{equation}
where $h_{IJ}(0)\equiv\left\langle \varphi_I(e)\varphi_J(0)\right\rangle,$ $C_{IJk}(0)\equiv\left\langle \varphi_I(\infty)\varphi_J(e){\cal O}_k(0)\right\rangle$, and for now we have dropped terms analytic in $\mu$. 

Assume now that a collection of operators $\varphi_I(x)$, $I=1,\cdots,L$ have constant dimension $\Delta$ on the conformal manifold. As $\lambda$ varies, these operators mix with one another. Provided we stay away from points where other operators acquire the same dimension $\Delta$, these operators form a $L$-dimensional vector bundle over the conformal manifold.  This bundle also has a natural connection and corresponding curvature. To see this, consider the counterterm
\begin{equation}
    \begin{split}
        \int \frac{d^d x}{S_{d-1}} A_i{}^{K}{}_L\lambda^i(x)a^L(x) {\varphi}_K(x)~,
    \end{split}
    \label{ctterms_2}
\end{equation}
 which contributes to \eqref{2pt_CPT_allops}. This counterterm can be interpreted as the contact term
\begin{equation}
    \begin{split}
        \left\langle \varphi_I(e)\varphi_J(0){\cal O}_i(x)\right\rangle=\delta^{(d)}(x) S_{d-1}A_i{}^K{}_J h_{IK}+\delta^{(d)}(x-e) S_{d-1}A_i{}^K{}_I h_{JK}~.
    \end{split}
    \label{ctct_term2}
\end{equation}
Note that there is no contribution from separated points since \eqref{2pt_CPT_allops} implies that $C_{IJk}$ vanishes for operators with constant scaling dimension. 

The contact term \eqref{ctct_term2} gives a connection for the vector bundle of operators $\varphi_I(x)$. We are free to choose counterterms so that this connection vanishes in the undeformed CFT. However, as in the case of exactly marginal operators \eqref{int_4pt}, upon deforming the theory this contact term will in general be produced. More precisely, we have
\begin{equation}
    \begin{split}
        \int \frac{d^d x_2}{S_{d-1}}\langle\varphi_I(e)\varphi_J(0){\cal O}_k(x_1){\cal O}_\ell(x_2)\rangle=S_{d-1}\partial_\ell\left(A_{k}{}^{K}{}_{J}h_{K I}(0)\right)\delta^{(d)}(x_1)+S_{d-1}\partial_\ell\left(A_{k}{}^{K}{}_{I}h_{KJ}(0)\right)\delta^{(d)}(x_1-e)~.
    \end{split}
    \label{int_4pt2}
\end{equation}
The LHS is defined using analytic regularization. The curvature of this vector bundle over the conformal manifold can be obtained from the RHS of \eqref{int_4pt2}, and is given by \eqref{Fintro}.

\section{CFT Curvature as an Inversion Formula}
\label{sec:inversion}

In the previous section, we have shown how the curvature of vector bundles on the conformal manifold, in particular the Riemann curvature of the tangent bundle, are computed from CFT 4-point functions. These expressions obscure the relation between the curvature and the CFT data, as they involves a double-integral of the regularized 4-point function.

In this section, we will relate these geometric properties of the conformal manifold to the inversion formula of CFT 4-point functions. This allows us to write the curvature of vector bundles as a particular value of a meromorphic function in dimension and spin, which has a clear meaning in terms of the spectrum of operators appearing in the OPE of the exactly marginal deformations $\mathcal{O}_{i}$. We further discuss the implication of the inversion formulas for points in the conformal manifold where curvatures diverge, and derive sum rules relating curvatures to the basic CFT data of scaling dimensions and OPE coefficients. 

Finally, we also discuss an inversion formula for the second derivative of the Zamolodchikov metric, and also derive associated sum rules.

\subsection{Euclidean Inversion Formula for the Curvature}

The curvature of the vector bundle of scalar operators $\varphi_I$ over the conformal manifold can be obtained from the expression
\begin{equation}
    \begin{split}
         F_{ij}{}^K{}_L(0)&=\int \frac{d^d x_1}{S_{d-1}}\int \frac{d^d x_2}{S_{d-1}}\left(x_{12}\cdot e\right)\langle\varphi^K(e)\varphi_L(0){\cal O}_i(x_1){\cal O}_j(x_2)\rangle_c~,
    \end{split}
    \label{Rijkl}
\end{equation}
To see this, we use \eqref{int_4pt2} to perform either the $x_1$ or $x_2$ integral first. After performing the second integral, the result is equal to \eqref{Fintro} in a scheme where $h_{IJ,i}=0$. The disconnected part of the correlation function does not enter in this expression, as discussed below \eqref{gij_dkdl}.

It is convenient to assume that only irrelevant operators appear in the OPE of the operators in \eqref{Rijkl}, in which case this equation does not need to be regularized. Contributions of relevant operators will be discussed in section \ref{sec:relops}. With this assumption, \eqref{Rijkl} equals:
\begin{equation}
    \begin{split}
        F_{ij}{}^K{}_L(0)=\frac{{\text {vol}}(SO(d-1))}{(S_{d-1})^2} I_{\Delta=1,J=1}~,
    \end{split}
    \label{Rijkl_EIF}
\end{equation}
where $I_{\Delta,J}$ is given by
\begin{equation}
    \begin{split}
        I_{\Delta,J}=\int \frac{d^d x_1d^d x_{2}}{{\text {vol}}(SO(d-1))}|x_{12}|^\Delta \hat C_{J}\left(\frac{x_{12}\cdot e}{|x_{12}|}\right)\langle\varphi^K(e)\varphi_L(0){\cal O}_i(x_1){\cal O}_j(x_2)\rangle_c~,
    \end{split}
    \label{EIF_2int}
\end{equation}
and $\hat C_J(x)$ is the Gegenbauer polynomial,
\begin{equation}
    \begin{split}
        &\hat C_J(x)=\frac{\Gamma(J+1)\Gamma\left(\frac{d-2}{2}\right)}{2^J\Gamma\left(J+\frac{d-2}{2}\right)}C_J^{\frac{d}{2}-1}(x)~,
        \\
        &C_J^{\frac{d}{2}-1}(x)=\frac{\Gamma(J+d-2)}{\Gamma(J+1)\Gamma(d-2)}{}_2F_1\left(-J,J+d-2,\frac{d-1}{2},\frac{1-x}{2}\right)~.
    \end{split}
\end{equation}
The expression \eqref{EIF_2int} is the Euclidean inversion formula for the (connected) 4-point function $\langle\varphi^K(e)\varphi_L(0){\cal O}_i(x_1){\cal O}_j(x_2)\rangle_c$ \cite{Caron-Huot:2017vep,Simmons-Duffin:2017nub}. It is valid for integer spin $J$.

In general, a CFT four-point function can be expanded as \cite{Caron-Huot:2017vep,Simmons-Duffin:2017nub}
\begin{equation}
    \begin{split}
        \left\langle \varphi_{I_1}(x_1)\cdots\varphi_{I_4}(x_4)\right\rangle=\int_{-\infty}^{+\infty}\frac{d\Delta}{2\pi i}\frac{I_{\Delta,J}}{n_{\Delta,J}}K_{\tilde\Delta,J}G_{\Delta,J}^{\Delta_{I_i}}(x_i)+(\text{non-norm})~,
    \end{split}
    \label{4ptdecomp}
\end{equation}
where $n_{\Delta,J}$ is a normalization factor that will not be important for us, and $G_{\Delta,J}^{\Delta_{I}}(x_i)$ is the $d$-dimensional conformal block, see \cite{Simmons-Duffin:2017nub} for conventions. The function $K_{\tilde\Delta,J}$ is given by
\begin{equation}
    \begin{split}
        K_{\tilde\Delta,J}=\left(-\frac{1}{2}\right)^J\pi^\frac{d}{2}\frac{\Gamma\left(\tilde\Delta-\frac{d}{2}\right)}{\Gamma\left(\tilde\Delta-1\right)}\frac{\Gamma\left(\tilde\Delta+J-1\right)}{\Gamma\left(d-\tilde\Delta+J\right)}\frac{\Gamma\left(\frac{d-\tilde\Delta+J}{2}\right)^2}{\Gamma\left(\frac{\tilde\Delta+J}{2}\right)^2}~,
    \end{split}
    \label{Kcoeff}
\end{equation}
and $I_{\Delta,J}$ is the coefficient function that encodes the CFT data in its analytic structure. More explicitly, the function $I_{\Delta,J}$ has simple-poles at the dimensions of primary operators exchanged in the $s$-channel OPE, with the residues related to structure constants. In this sense, it inverts the 4-point function. The non-normalizable contributions to \eqref{4ptdecomp} can come from the $s,t,u$-channels, and are related to relevant operators in these channels, which we will discuss below.

The Euclidean inversion formula can be written as a single integral over the cross-ratio $z,\bar z$,
\begin{equation}
    \begin{split}
        z \bar z=\frac{x^2_{12}x^2_{34}}{x^2_{13}x^2_{24}}~,~~~~(1-z)(1-\bar z)=\frac{x^2_{14}x^2_{23}}{x^2_{13}x^2_{24}}~.
    \end{split}
    \label{cr}
\end{equation}
We have \cite{Caron-Huot:2017vep,Simmons-Duffin:2017nub}:
\begin{equation}
    \begin{split}
        I_{\Delta,J}=\int \frac{d^2z}{2{\text {vol}}(SO(d-2))} |z|^{-2d+2\Delta_\varphi}&\left|\frac{z-\bar z}{2i}\right|^{d-2}  \langle\varphi^K(0)\varphi_L(z,\bar z){\cal O}_i(1){\cal O}_j(\infty)\rangle_c 
        \\        &\times\left[K_{\Delta,J}G_{\tilde\Delta,J}(z,\bar z)+K_{\tilde\Delta,J}G_{\Delta,J}(z,\bar z)\right]~,
    \end{split}
    \label{EIF_1int}
\end{equation}
where $\Delta_\varphi$ is the dimension of $\varphi_I$, $\tilde\Delta\equiv d-\Delta$, and the function $G_{\Delta,J}(z,\bar z)$ are the $d$-dimensional conformal block, normalized as
\begin{equation}
    \begin{split}
        G_{\Delta,J}(z,\bar z)\sim (z\bar z)^{\frac{\Delta}{2}}\left(\frac{z}{\bar z}\right)^{-\frac{J}{2}}~,~~~~ z\ll \bar z\ll1~.
    \end{split}
    \label{CBnorm}
\end{equation}

In $2d$, the conformal blocks $G_{\Delta,J}(z,\bar z)$ are given by
\begin{equation}
    \begin{split}
        G_{\Delta,J}(z,\bar z)=\frac{1}{1+\delta_{J,0}}\left(k_{\Delta+J}(z)k_{\Delta-J}(\bar z)+k_{\Delta+J}(\bar z)k_{\Delta-J}(z)\right)~,
    \end{split}
    \label{CB2d}
\end{equation}
where $k_{2h}(z)$ is the holomorphic $2d$ global conformal block,
\begin{equation}
    \begin{split}
        k_{2h}(z)=z^h {}_2F_1(h,h,2h,z)~.
    \end{split}
    \label{2dCB}
\end{equation}
Evaluating \eqref{Rijkl_EIF} for $d=2,\Delta=J=1$, we find
\begin{equation}
    \begin{split}
        F_{ij}{}^K{}_L(0)=\frac{1}{4\pi}\int d^2 z\ln|1-z|^2 |z|^{-4+2\Delta_\varphi}\langle\varphi^K(0)\varphi_L(z,\bar z){\cal O}_i(1){\cal O}_j(\infty)\rangle_c~.
    \end{split}
    \label{Rijkl_EIF1int_2d}
\end{equation}
The subscript $\langle(\cdots)\rangle_c$ denotes the connected part of the correlation function, \emph{i.e.}\ after subtracting the identity contribution in the $s,t,u$ channels. We will show this and also discuss the regularization of \eqref{Rijkl_EIF1int_2d} below, when we consider the contribution of relevant operators to \eqref{Rijkl_EIF1int_2d}.

In the case of the curvature of the tangent bundle (\emph{i.e.}\ the curvature of the Zamolodchikov metric), the expression \eqref{Rijkl_EIF1int_2d} was derived in \cite{Friedan:2012hi} by starting from an expression similar to \eqref{Rijkl}, which was obtained from the Weyl anomaly \cite{Osborn:1991gm,Friedan:2009ik}. Here we have seen that it follows directly from the Euclidean inversion formula for the 4-point function of exactly marginal operators.

In $4d$, we instead have
\begin{equation}
    \begin{split}
        G_{\Delta,J}(z,\bar z)=\frac{z \bar z}{\bar z- z}\left(k_{\Delta-J-2}(z)k_{\Delta+J}(\bar z)-k_{\Delta+J}( z)k_{\Delta-J-2}(\bar z)\right)~.
    \end{split}
    \label{4dCB}
\end{equation}
Evaluating \eqref{Rijkl_EIF} at $d=4,\Delta=J=1$, we obtain
\begin{equation}
    \begin{split}
        F_{ij}{}^K{}_L(0)=\frac{1}{2\pi}\int d^2z   |z|^{-8+2\Delta_\varphi}\left|\frac{z-\bar z}{2i}\right|^2\ln|1-z|^2\langle\varphi^K(0)\varphi_L(z,\bar z){\cal O}_i(1){\cal O}_j(\infty)\rangle_c~.
    \end{split}
    \label{Rijkl_EIF1int_4d}
\end{equation}

In Appendix \ref{sec:PWalld}, we show that
\begin{equation}
    \begin{split}
        K_{\Delta=1,J=1}G_{\tilde\Delta=d-1,J=1}(z,\bar z)+K_{\tilde\Delta=d-1,J=1}G_{\Delta=1,J=1}(z,\bar z)=\frac{1}{2}S_{d-1}\ln|1-z|^2~,
    \end{split}
    \label{PW11alld}
\end{equation}
for any dimension $d$. Using this in \eqref{Rijkl_EIF} and \eqref{EIF_1int}, we find an Euclidean inversion formula for the curvature $F_{ij}{}^K{}_L$ in any dimension $d$,
\begin{equation}
    \begin{split}
        F_{ij}{}^K{}_L(0)=A(d)\int d^2z |z|^{-2d+2\Delta_\varphi}   \left|\frac{z-\bar z}{2i}\right|^{d-2}\ln|1-z|^2\langle\varphi^K(0)\varphi_L(z,\bar z){\cal O}_i(1){\cal O}_j(\infty)\rangle_c ~,
    \end{split}
    \label{RijklEinvalld}
\end{equation}
where $A(d)=\frac{S_{d-2}}{4S_{d-1}}$.

In Appendix \ref{sec:bianchi} we check that the Bianchi identity for the Riemann curvature \eqref{RijklEinvalld} is satisfied.

\subsubsection{Relevant Operators}
\label{sec:relops}

Let us now comment on the contribution of relevant scalar operators to the OPE between the operators $\varphi_I,{\cal O}_i$. Their contribution to the Euclidean inversion formula is discussed in \cite{Simmons-Duffin:2017nub} for example (see also \cite{Caron-Huot:2017vep}). These operators spoil the normalizability of the 4-point function, and for the Euclidean inversion formula to be applicable, the contribution from relevant scalars needs to be subtracted from the 4-point function. In practice, what this means is that we introduce a hard-sphere cutoff $\epsilon$ in the integral near $z=0,1,\infty$, perform the integral for small, fixed $\epsilon>0$, and then drop divergent terms in $\epsilon$. 

We will now show that the hard-sphere and analytic regularizations give the same answer for the finite part of the correlation function. In the presence of relevant operators, the correlation function in \eqref{Rijkl} should be replaced by the regulated correlation function \eqref{reg_npt}. For a 4-point function, we have to replace
\begin{equation}
    \begin{split}
      \langle\varphi^K(e)\varphi_L(0){\cal O}_i(x_1){\cal O}_j(x_2)\rangle\to\mu^{4 s}\left(x^2_1x^2_2(1-x_1)^2(1-x_2)^2x^2_{12}\right)^{\frac{s}{3}}\langle\varphi^K(e)\varphi_L(0){\cal O}_i(x_1){\cal O}_j(x_2)\rangle~.
    \end{split}
    \label{reg_4pt}
\end{equation}
The $x_i$-dependent prefactor in \eqref{reg_4pt} can be written as
\begin{equation}
    \begin{split}
        x^2_1x^2_2(e-x_1)^2(e-x_2)^2x^2_{12}=(x^2_{12})^3\frac{(1-z)(1-\bar z)}{(z\bar z)^2}~,
    \end{split}
\end{equation}
where $z,\bar z$ are the cross-ratio given by \eqref{cr}. Thus, the regularized expression for $R_{ijk\ell}$ is
\begin{equation}
    \begin{split}
         F^{(s)}_{ij}{}^K{}_L(0)(0)&=\int \frac{d^d x_1}{S_{d-1}}\int \frac{d^d x_2}{S_{d-1}}\left(x_{12}\cdot  e\right)(x^2_{12})^s H_s(z,\bar z)~,
    \end{split}
    \label{Rijkl_reg}
\end{equation}
where
\begin{equation}
    \begin{split}
        H_s(z,\bar z)\equiv\mu^{4s}\left(\frac{(1-z)(1-\bar z)}{(z\bar z)^2}\right)^\frac{s}{3}\langle\varphi^K(e)\varphi_L(0){\cal O}_i(x_1){\cal O}_j(x_2)\rangle_c~.
    \end{split}
    \label{Hs}
\end{equation}
Thus, \eqref{Rijkl_reg} can be written in terms of \eqref{EIF_2int} as
\begin{equation}
    \begin{split}
         F^{(s)}_{ij}{}^K{}_L(0)(0)&=\frac{{\text {vol}}(SO(d-1))}{(S_{d-1})^2} I^{(s)}_{\Delta=1+2s,J=1}~,
    \end{split}
    \label{regRijkl_EIF_2int}
\end{equation}
where the superscript on $I_{\Delta,J}^{(s)}$ denotes that the inversion formula is applied on the regulated correlation function $H_s(z,\bar z)$ given by \eqref{Hs}, \IE
\begin{equation}
    \begin{split}
        I^{(s)}_{\Delta,J}=\int \frac{d^2z}{2{\text {vol}}(SO(d-2))}|z|^{-2d+2\Delta_\varphi}\left|\frac{z-\bar z}{2i}\right|^{d-2}   H_s(z,\bar z) \left[K_{\Delta,J}G_{\tilde\Delta,J}(z,\bar z)+K_{\tilde\Delta,J}G_{\Delta,J}(z,\bar z)\right]~.
    \end{split}
    \label{regEIF_1int}
\end{equation}
The integrand in \eqref{regEIF_1int} is finite at $s=0$ away from the singularities at $z=0,1,\infty$. Thus, poles in $s$ can only come from the integral near these points. 

Consider now the contribution of a relevant scalar operator of dimension $\Delta$ in the $t$-channel. We only need to keep the $s$ dependence that will regulate the singularity near $z=0,1,\infty$. Then we see that the singularity near $z=1$ is regulated by the $s$-dependent prefactor in \eqref{Hs}, \IE the contribution of the relevant operator to \eqref{regEIF_1int} is given by 
\begin{equation}
    \begin{split}
        F^{(s)}_{ij}{}^K{}_L(0)&\supset A(d)\mu^{4s}\int_{|1-z|<\epsilon_2} d^2 z  |z|^{-2d+2\Delta_\varphi}\left|\frac{z-\bar z}{2i}\right|^{d-2}  |1-z|^\frac{2s}{3}C_{Li}{}^AC^K{}_{jA}\frac{1}{|1-z|^{d+\Delta_\varphi-\Delta}}\ln|1-z|^2~,
        \\
        &\to C_{Li}{}^AC^K{}_{jA}\epsilon_2^{\Delta-\Delta_\varphi}\frac{-1+(\Delta-\Delta_\varphi)\ln\epsilon_2}{(\Delta-\Delta_\varphi)^2}~,
    \end{split}
    \label{regEIF_relt}
\end{equation}
where in the second line we took the limit $s\to0$ after performing the integral. From below \eqref{ctct_term2}, $C_{Li}{}^A=C^K{}_{jA}=0$ if $\Delta=\Delta_\varphi$, so there is no divergence at this value of $\Delta$. The same contribution in the hard-sphere regularization is
\begin{equation}
    \begin{split}
        &A(d)\int_{\epsilon_1<|1-z|<\epsilon_2} d^2 z  \left|\frac{z-\bar z}{2i}\right|^{d-2}  C_{Li}{}^AC^K{}_{jA}\frac{1}{|1-z|^{d+\Delta_\varphi-\Delta}}\ln|1-z|^2
        \\
        &~~~~\to C_{Li}{}^AC^K{}_{jA}\epsilon_2^{\Delta-\Delta_\varphi}\frac{-1+(\Delta-\Delta_\varphi)\ln\epsilon_2}{(\Delta-\Delta_\varphi)^2}~,
    \end{split}
\end{equation}
where in the second line we dropped terms that diverge with $\epsilon_1$. Thus, we see that analytic regularization gives the same answer as hard-sphere regularization of \eqref{EIF_1int}. 

In the Euclidean inversion formula, the hard-sphere cutoff is used to regularize the contribution from relevant scalars \cite{Caron-Huot:2017vep,Simmons-Duffin:2017nub}. The argument above shows that this agrees with analytic regularization, and hence we have established the result \eqref{Rijkl_EIF} also in the presence of relevant operators, where \eqref{EIF_1int} is to be regularized with the hard-sphere cutoff, or using analytic regularization as in \eqref{regEIF_1int}.

Starting from \eqref{regEIF_1int} and \eqref{Hs}, we can now consider the vacuum contribution in the $t,u$ channels. In the analytic regularization scheme, this evaluates to zero. This explains why we have dropped these contributions in the correlation functions \eqref{gij_dkdl} and \eqref{Rijkl}. In the case of exactly marginal deformations, the vanishing of these contributions is also expected from the Weyl anomaly \cite{Friedan:2012hi,Gomis:2015yaa}.

\subsection{Conformal Manifolds with Infinite Curvature}

It is interesting to explore what \eqref{Rijkl_EIF} implies for conformal manifolds where the curvature diverges. Since the kernel in \eqref{EIF_1int} is invariant under $\Delta\leftrightarrow d-\Delta$, it follows that the function $I_{\Delta,J}$ also respects this symmetry \cite{Caron-Huot:2017vep,Simmons-Duffin:2017nub}. Thus, \eqref{EIF_2int} gives a relation between the curvature $F_{ij}{}^K{}_L$ and the coefficient function $I_{\Delta=d-1,J=1}$. We note in particular that these are the quantum numbers of a conserved current operator.

The analytic structure of the function $I_{\Delta,J}$ implies that, for integer $J\geq0$ and $\Delta\geq d/2$, it diverges only at the location of physical operators in the $s$-channel OPE. Thus, if the curvature $F_{ij}{}^K{}_L$ diverges and the analytic structure of $I_{\Delta,J}$ is preserved, it follows that there is an operator of dimension $\Delta=d-1$ and $J=1$ in the $s$-channel OPE. 

It is possible that the analytic structure of $I_{\Delta,J}$ is modified as the theory is deformed to the point where the curvature diverges. For example, the spectrum of the CFT can become continuous in this limit. This happens for example in the conifold singularity of $(2,2)$ SCFTs on Calabi-Yau manifolds in $2d$ \cite{Candelas:1990rm}, where the Riemann curvature diverges as the target space of the theory becomes non-compact.

In summary, we find that the curvature of a vector bundle over the conformal manifold can diverge only if the theory develops a continuum, or if a conserved current appears in the OPE of exactly marginal operators ${\cal O}_i$ and operator $\varphi_I$. For two-dimensional CFTs, a pole at $\Delta=1$ is not invariant under the shadow transformation $\Delta\to 2-\Delta$, and hence the only way the curvature can diverge is if the theory develops a continuum.

Note that we are assuming the divergence in the curvature is not delta-function localized on the conformal manifold, as in the case of orbifold singularities (for example points of enhanced symmetry on Narain moduli spaces).

\subsection{Lorentzian Inversion Formula for the Curvature}

The coefficient function $I_{\Delta,J}$ also admits an expression as a double commutator of a 4-point function integrated over a Lorentzian regime \cite{Caron-Huot:2017vep, Simmons-Duffin:2017nub}. This expression satisfies several nice properties. It is analytic in both dimension $\Delta$ and spin $J$, and satisfies positivity properties. The latter is due to the fact that the double-discontinuity entering in the integral of $I_{\Delta,J}$ is a positive-definite function when the first two and last two operators are pairwise identical.

The Lorentzian inversion formula is given by \cite{Caron-Huot:2017vep, Simmons-Duffin:2017nub}
\begin{equation}
    \begin{split}
I_{\Delta,J}=\alpha_{\Delta,J}\Bigg[\int_0^1dz\int_0^1 d\bar z &|z|^{-2d+2\Delta_\varphi}|z-\bar z|^{d-2} G_{J+d-1,\Delta-d+1}(z,\bar z)
        \\
        &\times\Big((-1)^J\left\langle[{\cal O}_i,\varphi_L][\varphi^K,{\cal O}_j]\right\rangle_c+\left\langle[{\cal O}_j,\varphi_L][\varphi^K,{\cal O}_i]\right\rangle_c\Big)\Bigg]~,
    \end{split}
    \label{LIF_1int}
\end{equation}
where $z,\bar z$ are independent real variables. The function $\alpha_{\Delta,J}$ is defined as
\begin{equation}
    \begin{split}
        &\alpha_{\Delta,J}=-\frac{a_{\Delta,J}}{2^d}\frac{\hat C_J(1)}{{\text{vol}}SO(d-1)}~,
        \\
        &a_{\Delta,J}=\frac{1}{2}(2\pi)^{d-2}\frac{\Gamma(J+1)}{\Gamma\left(J+\frac{d}{2}\right)}\frac{\Gamma\left(\Delta-\frac{d}{2}\right)}{\Gamma\left(\Delta-1\right)}\frac{\Gamma\left(\frac{J+\Delta}{2}\right)^2\Gamma\left(\frac{J+d-\Delta}{2}\right)^2}{\Gamma(J+\Delta)\Gamma(J+d-\Delta)}~,
    \end{split}
\end{equation}
and the double-discontinuity $\left\langle[{\cal O}_i,\varphi_L][\varphi^K,{\cal O}_j]\right\rangle_c$ in \eqref{LIF_1int} is defined as follows. Start from the 4-point function
\begin{equation}
    \begin{split}
        \left\langle \varphi^K(0)\varphi_L(z,\bar z){\cal O}_i(1){\cal O}_j(\infty)\right\rangle=\frac{1}{|z|^{2\Delta_\varphi}}g(z,\bar z)~.
    \end{split}
    \label{gdef}
\end{equation}
Then
\begin{equation}
    \begin{split}
        \left\langle[{\cal O}_i,\varphi_L][\varphi^K,{\cal O}_j]\right\rangle=-\frac{2}{|z|^{2\Delta_\varphi}}{\rm dDisc}\left[g(z,\bar z)\right]~,
    \end{split}
    \label{dDisc1}
\end{equation}
where
\begin{equation}
    \begin{split}
        {\rm dDisc}\left[g(z,\bar z)\right]=g(z,\bar{z})-\frac{1}{2}g^{\circlearrowleft}(z,\bar{z})-\frac{1}{2}g^{\circlearrowright}(z,\bar{z})~,
    \end{split}
    \label{dDisc2}
\end{equation}
where $g^{\circlearrowleft}(z,\bar{z})$ denotes starting from the $g(z,\bar{z})$ with $0<\bar{z},z<1$, going around $\bar{z}=1$ counterclockwise, and returning to the original value of $\bar{z}$. $g^{\circlearrowright}(z,\bar{z})$ is analagous, but we move $\bar z$ around $\bar z=1$ in a clockwise direction. The double-discontinuity $\left\langle[{\cal O}_i,\varphi_L][\varphi^K,{\cal O}_j]\right\rangle_c$ is defined from \eqref{dDisc1} by subtracting the $t,u$-channel contributions from the vacuum (the $s$-channel contribution from the vacuum vanishes).

When the operators are pairwise equal, \IE $\varphi^K=\varphi_L$  and ${\cal O}_i={\cal O}_j$, then ${\rm dDisc}\left[g(z,\bar z)\right]\geq 0$. Note that while this is true of ${\rm dDisc}\left[g(z,\bar z)\right]\geq 0$, it is not necessarily true that ${\rm dDisc}\left[g_c(z,\bar z)\right]\geq 0$, since the vacuum exchange in $t,u$ channels contribute to ${\rm dDisc}\left[g(z,\bar z)\right]$.

Once subtlety in obtaining the Lorentzian inversion formula \eqref{LIF_1int} is that, for general operators, it is equal to the Euclidean inversion formula only for $J>1$. In Appendix \ref{sec:LIFsubtleties}, we point out that for the 4-point function $\left\langle\varphi^K(0)\varphi_L(z,\bar z){\cal O}_i(1){\cal O}_j(\infty)\right\rangle$, we can still use the Lorentzian inversion formula to calculate $F_{ij}{}^K{}_L$. We also discuss there how to include relevant operators in the OPE of the operators $\varphi_I,{\cal O}_i$.

\subsection{Sum Rules for the Curvature}

The Lorentzian inversion formula \eqref{LIF_1int} allows us to write an expression for the curvature of the conformal manifold in terms of CFT data. In this section, we do this explicitly in $2d$ and $4d$, where the conformal blocks are known explicitly, see \eqref{CB2d}, \eqref{4dCB}. 

\subsubsection{$2d$}
 Expanding $\left\langle\varphi^K(0)\varphi_L(z,\bar z){\cal O}_i(1){\cal O}_j(\infty)\right\rangle$ in the $t$-channel gives
\begin{equation}
    \begin{split}
    \left\langle\varphi^K(0)\varphi_L(z,\bar z){\cal O}_i(1){\cal O}_j(\infty)\right\rangle=\sum_{A}C_{Li}{}^AC^K{}_{j A}(1-z)^{-1-\frac{\Delta_\varphi}{2}+h_A}(1-\bar z)^{-1-\frac{\Delta_\varphi}{2}+\bar h_A}~,
    \end{split}
    \label{4pt_tch}
\end{equation}
where the index $A$ runs over all operators $\varphi_A$ in the OPE of the operators $\varphi_L,{\cal O}_i$, with conformal weights $(h_A,\bar h_A)$. We take the conformal weights of the scalar operators $\varphi_I$ to be $(h_I, \bar h_I)=\left(\frac{\Delta_\varphi}{2},\frac{\Delta_\varphi}{2}\right)$. From this we find 
\begin{equation}
    \begin{split}
       \left\langle[{\cal O}_i,\varphi_L][\varphi^K,{\cal O}_j]\right\rangle=-2\sum_{A}C_{Li}{}^AC^K{}_{j A}(1-z)^{-1-\frac{\Delta_\varphi}{2}+h_A}(1-\bar z)^{-1-\frac{\Delta_\varphi}{2}+\bar h_A}\left[1-\cos\left(2\pi \left(\bar{h}_A- \frac{\Delta_\varphi}{2}\right)\right)\right]~.
    \end{split}
    \label{ijkl_tch}
\end{equation}
Similarly, 
\begin{equation}
    \begin{split}
        \left\langle[{\cal O}_j,\varphi_L][\varphi^K,{\cal O}_i]\right\rangle=-2\sum_{A}C_{Lj}{}^AC^K{}_{i A}(1-z)^{-1-\frac{\Delta_\varphi}{2}+h_A}(1-\bar z)^{-1-\frac{\Delta_\varphi}{2}+\bar h_A}\left[1-\cos\left(2\pi \left(\bar{h}_A-\frac{\Delta_\varphi}{2}\right)\right)\right]~.
    \end{split}
     \label{ijkl_uch}
\end{equation}
Using this in \eqref{LIF_1int}, we find
\begin{equation}
    \begin{split}
        F_{ij}{}^K{}_L&=\frac{1}{8\pi^2}\sum_{A}\left(C_{Li}{}^AC^K{}_{j A}-C_{ji}{}^AC^K{}_{i A}\right)\left[1-\cos\left(2\pi \left(\bar{h}_A-\frac{\Delta_\varphi}{2}\right)\right)\right]r_{\Delta_\phi}(h_A,\bar h_A)~,
    \end{split}
    \label{2dsumrule_all}
\end{equation}
where
\begin{equation}
    \begin{split}
        r_{\Delta_\varphi}(h,\bar h)&\equiv\frac{\Gamma(\Delta_\varphi-1)\Gamma\left(h-\frac{\Delta_\varphi}{2}\right)}{\Gamma\left(h+\frac{\Delta_\varphi}{2}-1\right)}\left(\psi^{(0)}\left(-\frac{\Delta_\varphi}{2}+h\right)-\psi^{(0)}\left(-1+\frac{\Delta_\varphi}{2}+h\right)\right)
        \\
 &\times\frac{\Gamma(\Delta_\varphi-1)\Gamma\left(\bar h-\frac{\Delta_\varphi}{2}\right)}{\Gamma\left(\bar h+\frac{\Delta_\varphi}{2}-1\right)}\left(\psi^{(0)}\left(-\frac{\Delta_\varphi}{2}+\bar h\right)-\psi^{(0)}\left(-1+\frac{\Delta_\varphi}{2}+\bar h\right)\right)~,
    \end{split}
\end{equation}
where $\psi^{(m)}(z)$ is the polygamma function of order $m$. The sum is over all operators in the OPE, both primaries and descendants, except for the identity contribution in the $t$-channel and $u$-channels. The $s$-channel contribution vanishes upon taking the double-discontinuity, so it need not be subtracted off.

Note that for $h_A<\frac{\Delta_\varphi}{2}$ or $\bar h_A<\frac{\Delta_\varphi}{2}$, the integral over $z,\bar z$ in \eqref{LIF_1int} is divergent. We regulate this by cutting a small region of size $\epsilon$ near $z,\bar z=1$, and dropping divergences in $\epsilon$. Equivalently, we can perform the integrals assuming $h_A>\frac{\Delta_\varphi}{2},\bar h_A>\frac{\Delta_\varphi}{2}$, and then analytically continue the result to the region $h_A<\frac{\Delta_\varphi}{2},\bar h_A<\frac{\Delta_\varphi}{2}$.

To arrive at \eqref{2dsumrule_all}, we  swapped the OPE sum and integration over $z,\bar z$. Let's discuss when this is allowed, focusing on the $t$-channel expansion in \eqref{ijkl_tch}. Exchanging the two limits is allowed provided that there are no singularities coming from the small $z,\bar z$ integral in \eqref{LIF_1int}. For $d=2$, the integrand of \eqref{LIF_1int} behaves as 
\begin{equation}
    \begin{split}
        |z|^{-4+2\Delta_\varphi}G_{2,0}(z,\bar z)\left\langle[{\cal O}_k,{\cal O}_j][{\cal O}_i,{\cal O}_\ell]\right\rangle\sim\ln(1-z)\ln(1-\bar z)z^{-2+h_s}\bar z^{-2+\bar h_s}~,
    \end{split}
\end{equation}
where $h_s,\bar h_s$ are the conformal weights of an operator in the $s$-channel OPE between the operators ${\cal O}_i$ and  ${\cal O}_j$. The contribution from the small $z,\bar z$ region to the integral in \eqref{LIF_1int} is finite provided that $h_s>0$ and $\bar h_s>0$. 

Thus, we find that the sum rule \eqref{2dsumrule_all} converges provided that there are no holomorphic conserved currents in the $s$-channel OPE. Note that the vacuum block does not appear in the $s$-channel OPE when $i\neq j$, and when $i=j$ it cancels between the $t,u$-channel contributions in \eqref{LIF_1int}.

In case there are holomorphic conserved currents present (as will be the case for free theories, for example), a simple solution is to subtract their contribution explicitly, and apply the sum rule above to the remaining operators. We will do an explicit example of this in section \ref{sec:examples}.

From \eqref{2dsumrule_all} we find the curvature of the tangent bundle, \IE the Riemannian curvature of the Zamolodchikov metric, which is given by
\begin{equation}
    \begin{split}
        R_{ijk\ell}&=\frac{1}{8\pi^2}\sum_{A}\left(C_{\ell i}{}^AC_{kj A}-C_{\ell j}{}^AC_{ki A}\right)\frac{\left[1-\cos\left(2\pi \bar{h}_A\right)\right]}{(1-h_A)^2(1-\bar h_A)^2}~.
    \end{split}
    \label{zam2dsumrule_all}
\end{equation}

We can also write a sum rule for global conformal primaries. For simplicity we will write it only for the curvature of the tangent bundle. Expanding \eqref{4pt_tch} in terms of global primaries, we find
\begin{equation}
    \begin{split}
        \left\langle\left[  {\cal O}_k,{\cal O}_j\right]\left[  {\cal O}_i,{\cal O}_\ell\right]\right\rangle=-2\sum_{B}C_{jk}{}^BC_{i\ell B}(1-z)^{-2+h_B}(1-\bar z)^{-2+\bar h_B}\left(1-\cos(2\pi \bar{h}_B)\right) k_{2h_B}(z)\tilde k_{2\bar h_B}(\bar z)~,
    \end{split}
\end{equation}
where the sum over $B$ runs over primaries with respect to the global conformal group, and $k_\beta(z)$ is the global conformal block \eqref{2dCB}. Using this in \eqref{LIF_1int}, we find
\begin{equation}
    \begin{split}
        R_{ijk\ell}=\frac{1}{8\pi^2}\sum_{B}\left(C_{j\ell }{}^BC_{i kB}-C_{jk}{}^BC_{i\ell B}\right)\frac{\Gamma(2h_B)}{\Gamma(h_B)^2}\frac{\Gamma(2\bar h_B)}{\Gamma(\bar h_B)^2}\frac{\left(1-\cos(2\pi \bar{h}_B)\right)}{(h_B-1)^2(\bar h_B-1)^2}~.
    \end{split}
\end{equation}
This sum rule converges as long as there are no higher spin conserved currents in the $s$-channel OPE.

\subsubsection{$4d$}

We can also obtain a sum rule for the curvature of $4d$ CFTs. For simplicity, we will focus on the curvature of the tangent bundle. The OPE expansion in the $t$-channel is
\begin{equation}
    \begin{split}        \langle{\cal O}_i(0){\cal O}_j(z,\bar z){\cal O}_{k}(1){\cal O}_{\ell}(\infty)\rangle_c=\sum_{A}C_{jk}{}^AC_{i\ell A} (1-z)^{-4+\frac{\Delta_A-J_A}{2}}(1-\bar z)^{-4+\frac{\Delta_A+J_A}{2}}~,
    \end{split}
\end{equation}
where the sum over the index $A$ is over all operators $V_A$ in the $t$-channel OPE, not necessarily global primaries. The double-discontinuity is given by
\begin{equation}
    \begin{split}
         \left\langle\left[  {\cal O}_k,{\cal O}_j\right]\left[  {\cal O}_i,{\cal O}_\ell\right]\right\rangle_c=-2\sum_{A}C_{jk}{}^AC_{i\ell A}&(1-z)^{-4+\frac{\Delta_A-J_A}{2}}(1-\bar z)^{-4+\frac{\Delta_A+J_A}{2}}
        \\
        &~~~\times\left[1-\cos\left(\pi\left(\Delta_A+J_A\right)\right)\right]~.
    \end{split}
    \label{dDisc4d_tch}
\end{equation}
There is a similar expression for the $u$-channel double-discontinuity. Using this in \eqref{Rijkl_EIF},\eqref{LIF_1int}, we find
\begin{equation}
    \begin{split}
       R_{ijk\ell}&=\frac{32}{\pi^2}\sum_{A}\left(C_{jk A}C_{i\ell A}-C_{j\ell A}C_{ik A}\right)\left[1-\cos\left(\pi(\Delta_A+J_A)\right)\right]
        \\
        &\times\frac{7J_A^4-2(44+3(\Delta_A-8)\Delta_A)J_A^2-(\Delta_A-6)^2(\Delta_A-2)^2}{\left[(\Delta_A-2)^2-J_A^2\right]^2\left[(\Delta_A-4)^2-J_A^2\right]^2\left[(\Delta_A-6)^2-J_A^2\right]^2}~.
    \end{split}
    \label{4dsumrule_all}
\end{equation}
It is also possible to obtain a sum rule over conformal primaries only, though for simplicity we omit the explicit expression.

Convergence of the sum \eqref{4dsumrule_all} requires that all operators in the $s$-channel OPE have dimensions $\Delta_s>2+J_s$ (the sum rules for the curvature of other vector bundles also converge if this condition is satisfied). Operators that saturate this bound are conserved currents, and they lead to a divegence in the small $z,\bar z$ region of the $z,\bar z$ integral in \eqref{LIF_1int}. Note that  the contribution of finitely many such operators vanishes since the double-discontinuity of each vanishes individually. So the sum rule converges provided there are not infinitely many higher spin conserved currents in the $s$-channel OPE. It is known that the presence of these higher spin conserved current implies that the theory is free \cite{Maldacena:2011jn, Alba:2013yda, Alba:2015upa}.

\subsection{Inversion Formula for $g_{ij,k\ell}$}

So far, we discussed how to compute the curvature of vector bundles over the conformal manifold using CFT data. In this subsection, we will discuss an inversion formula for the second derivative of the metric, $g_{ij,k\ell}$. 

Unlike the Riemann curvature at the reference CFT,
$g_{ij,k\ell}$ is sensitive to the choice of local coordinates, \IE to the choice of counterterms. We assume we are working in the minimal subtraction scheme where the counterterms cancel all $1/s$ poles in analytic regularization, and have no finite part. Equations \eqref{gij_dkdl} and \eqref{EIF_2int} suggest that $g_{ij,k\ell}$ is related to $I_{\Delta=0,J=0}$, in local coordinates where $g_{ij,k}=0$. The goal of this section is to make this correspondence more precise. As we will see, the relation between $g_{ij,k\ell}$ and $I_{\Delta=0,J=0}$ is slightly more subtle than that between $R_{ijk\ell}$ and $I_{\Delta=1,J=1}$.

For simplicity, let's assume there are no relevant operators in the OPE between exactly marginal operators. Then \eqref{gij_dkdl} and \eqref{EIF_2int} suggest that
\begin{equation}
    \begin{split}
        g_{ij,k\ell}\stackrel{?}{=}\frac{{\text {vol}}(SO(d-1))}{(S_{d-1})^2} I_{\Delta=0,J=0}~.
    \end{split}
    \label{gijkl_naive}
\end{equation}
As we will see shortly, this is not quite correct as written.  In particular, in appendix \ref{sec:PWalld} we show that in any dimension $d$, we have
\begin{equation}
    \begin{split}
        K_{\Delta,J}G_{\tilde\Delta,J}(z,\bar z)+K_{\tilde\Delta,J}G_{\Delta,J}(z,\bar z)=2S_{d-1}\left(\frac{1}{\Delta}+\frac{1}{4}\ln\left(\frac{|z|^4}{|1-z|^2}\right)\right)+{\cal O}(\Delta)~.
    \end{split}
    \label{PW00alld}
\end{equation}
The leading contribution has a pole as $\Delta\to0$. The coefficient of this pole is proportional to \eqref{int_4pt_sep}, so it gives a $0\times\infty$ ambiguity to $g_{ij,k\ell}$.

To properly deal with this ambiguity, we must work with the regularized expression for $g_{ij,k\ell}$. Following the same steps that led to \eqref{regRijkl_EIF_2int}, we find that
\begin{equation}
    \begin{split}
        g^{(s)}_{ij,k\ell}=\frac{{\text {vol}}(SO(d-1))}{(S_{d-1})^2} I^{(s)}_{\Delta=2s,J=0}~,
    \end{split}
\end{equation}
where $I^{(s)}_{\Delta,J=0}$ is given by
\eqref{regEIF_1int}. As $s\to0$, we find
\begin{equation}
    \begin{split}
         &H_s(z,\bar z) \left[K_{\Delta=2s,J=0}G_{\tilde\Delta=d-2s,J=0}(z,\bar z)+K_{\tilde\Delta=d-2s,J=0}G_{\Delta=2s,J=0}(z,\bar z)\right]
         \\
         &~~~=S_{d-1}\left(\frac{1}{s}+4\ln\mu+\frac{1}{6}\ln\left(\frac{z^2\bar z^2}{(1-z)(1-\bar z)}\right)\right)\langle{\cal O}_i(1){\cal O}_j(0){\cal O}_k(x_1){\cal O}_\ell(x_2)\rangle+{\cal O}(s)~.
    \end{split}
    \label{gijkl_intgd}
\end{equation}
The coefficient of the pole in $s$ and the $\ln \mu$ term are the same, and is given by \eqref{int_4pt_sep}. This coefficient has to vanish since it contributes to the beta-function of the three-point function of exactly marginal operators. The remaining piece is finite in $s$. Comparing the finite piece in \eqref{gijkl_intgd} to \eqref{RikljEinv} and \eqref{RiljkEinv}, we find
\begin{equation}
    \begin{split}
        g_{ij,k\ell}=-\frac{1}{3}\left(R_{ikj\ell}+R_{i\ell jk}\right)~.
    \end{split}
    \label{gijkl_4}
\end{equation}
The final result \eqref{gijkl_4} is the expression for the second-derivative of the metric in Riemann normal coordinates. Thus, the subtraction scheme we have been using so far where all counterterms are set to zero is equivalent to working in Riemann normal coordinates.

To obtain the result \eqref{gijkl_4}, it was important to keep track of the regulator, due to the $0\times\infty$ ambiguity in the ``naive" inversion formula \eqref{gijkl_naive}. Had we used \eqref{gijkl_naive} directly, we would have found the incorrect result
\begin{equation}
    \begin{split}
        g_{ij,k\ell}\stackrel{?}{=}-\left(R_{ikj\ell}+R_{i\ell jk}\right)~,
    \end{split}
    \label{gijklRNCnaive}
\end{equation}
where we kept only the finite piece in \eqref{PW00alld} and dropped the pole in $\Delta$. Comparing against \eqref{gijkl_4}, we see that they differ only in the overall numerical factor. Hence, when discussing the inversion formula for $g_{ij,k\ell}$, it is easier to work with the ``naive" inversion formula \eqref{gijkl_naive}, drop the pole $\frac{1}{\Delta}$ that arises before performing the integral, perform the integral, and then divide the result by 3. As we have shown, this is equal to the regularized expression, and allows us to work with the analytically continued inversion formula $I_{\Delta,J}$. In other words, we have
\begin{equation}
    \begin{split}
        g_{ij,k\ell}=\frac{1}{3}\frac{{\text {vol}}(SO(d-1))}{(S_{d-1})^2}\left.\lim_{\Delta\to0} I_{\Delta,J=0}\right|_{\text{finite}}~,
    \end{split}
    \label{gijkl_EIF}
\end{equation}
where we only keep the finite piece of $I_{\Delta,J=0}$ in the limit $\Delta\to0$. In this way, we see that the function $I_{\Delta,J}$ has the property that both $I_{\Delta=0,J=0}$ and $I_{\Delta=1,J=1}$ are related to the Riemann curvature of the conformal manifold.

\subsection{Sum Rules for the Second Derivative of the Metric}

The inversion formula \eqref{gijkl_EIF} gives a sum rule for $g_{ij,k\ell}$. Since we are working in Riemann normal coordinates, this can in turn be related to a sum rule for the Riemann curvature in these coordinates, which we now discuss in $2$ and $4$ dimensions.

\subsubsection{$2d$}

Using \eqref{LIF_1int}, we find
\begin{equation}
    \begin{split}
        g_{ij,k\ell}=-\lim_{\delta\to0}\frac{1}{24\pi^2\delta}\int_0^1 dz d\bar z G_{\Delta=1,J=\delta-1}(z,\bar z)\left(\left\langle[{\cal O}_k,{\cal O}_j][{\cal O}_i,{\cal O}_\ell]\right\rangle+(k\leftrightarrow\ell)\right).
    \end{split}
    \label{gijklsumrule1}
\end{equation}
Using \eqref{CB2d}, we find that the block entering in the kernel of \eqref{gijklsumrule1} admits the small $\delta$ expansion:
\begin{equation}
    \begin{split}
        G_{\Delta=1,J=\delta-1}(z,\bar z)&=-\ln|1-z|^2-\frac{\delta}{2}\left[\ln z\ln(1-\bar z)+\ln \bar z\ln(1- z)-\ln(1-z)\ln(1-\bar z)\right]
        \\
        &+\frac{\delta}{2}\left[\ln z \ln(1-z)+\ln \bar z \ln(1-\bar z)+2\ln|1-z|^2+2{\text{Li}}_2(z)+2{\text{Li}}_2(\bar z)\right]~,
    \end{split}
    \label{CB2dgijkl}
\end{equation}
where we included terms up to order ${\cal O}(\delta)$. 

The double-commutators entering in the inversion formula are given by \eqref{ijkl_tch} and \eqref{ijkl_uch}. The leading term in \eqref{CB2dgijkl} contributes to a pole $1/\delta$ on the RHS of \eqref{gijklsumrule1}. As discussed below \eqref{gijkl_intgd}, this pole has to vanish by exact marginality. Hence, assuming that we can commute the sum over operators with the integrals over $z,\bar z$ (we will soon see that this is not quite right), we find the sum rule
\begin{equation}
    \begin{split}
        \frac{1}{12\pi^2}\sum_{A}\left(C_{jkA}C_{i\ell A}+C_{j\ell A}C_{i k A}\right)\left(1-\cos(2\pi \bar{h}_A)\right)\frac{h_A+\bar h_A-2}{(h_A-1)^2(\bar h_A-1)^2}=0~,
    \end{split}
    \label{zerosumrule}
\end{equation}
where the sum is over all operators in the theory (primaries and descendants). 

As in the previous sum rule \eqref{2dsumrule_all}, we have to worry about convergence of the $z,\bar z$ integral in \eqref{gijklsumrule1} near $z=0$ or $\bar z=0$, where the $t$-channel expansion does not converge. From the kernel \eqref{CB2dgijkl}, we see that the integral over $z$ has a divergence near $z=0$ if the double-discontinuity behaves as $z^{-2+h_s}$ near $z=0$, for any\footnote{Operators with weights $(1,n)$ or $(n,1)$, $n\geq1$, do not appear in the OPE of exactly marginal operators. The argument for this is a simple generalization of that leading to \eqref{Oi_anomdim}, allowing for operators with non-zero spin. See \cite{Kutasov:1988xb} for more details.} $0\leq h_s< 1$. Thus, a sufficient condition for convergence of the integral is that there are no operators with weights $(h_s,\bar h_s)$ satisfying $0\leq h_s<1$ or $0\leq\bar h_s< 1$ in the $s$-channel OPE.  

From now on we assume this condition to be true, the only exception being the vacuum block when the operators are identical. The sum rule \eqref{zerosumrule} is then modified to
\begin{equation}
    \begin{split}
        \frac{1}{12\pi^2}\sum_{A}\left(C_{jkA}C_{i\ell A}+C_{j\ell A}C_{i k A}\right)\left(1-\cos(2\pi \bar{h}_A)\right)\frac{h_A+\bar h_A-2}{(h_A-1)^2(\bar h_A-1)^2}+V_0 \delta_{ij}\delta_{k\ell}=0~,
    \end{split}
    \label{zerosumrulevac}
\end{equation}
where $V_0$ is the contribution from the double-discontinuity of the vacuum Virasoro block, and on the LHS we must subtract this contribution from the $t$-channel sum. Of course, since the vacuum Virasoro block is not known in closed form, we do not know what $V_0$ is, but in any case it should be a constant (depending only on the central charge $c$). At large $c$, the Virasoro block is equal to the global block, and therefore $V_0\to0$ as $c\to\infty$. We do not know of an expression for $V_0$ at finite $c$. 

Let's consider the case $i=j\neq k=\ell$, so that \eqref{zerosumrule} simplifies to
\begin{equation}
    \begin{split}
        \frac{1}{6\pi^2}\sum_{A}C_{ikA}^2\left(1-\cos(2\pi \bar{h}_A)\right)\frac{h_A+\bar h_A-2}{(h_A-1)^2(\bar h_A-1)^2}+V_0=0~,
    \end{split}
    \label{leadingsumrule_iikk}
\end{equation}
Since $C_{ikA}^2\left(1-\cos(2\pi \bar{h}_A)\right)\geq 0$, this sum rule implies that irrelevant operators in the OPE of the exactly marginal operators ${\cal O}_i$ and ${\cal O}_k$ contribute with a plus sign to the sum on the LHS, and relevant operators contribute with a negative sign to the sum on the LHS. Note that there can still be relevant operators in the OPE of these exactly marginal operators, so long as they do not appear in the $s$-channel OPE we are considering. 

At next order in $\delta$ in the expansion \eqref{CB2dgijkl}, we find the sum rule for $g_{ij,k\ell}$,
\begin{equation}
    \begin{split}
        g_{ij,k\ell}=\frac{1}{24\pi^2}\sum_A\left(C_{jkA}C_{i\ell A}+C_{j\ell A}C_{i k A}\right)\frac{\left(1-\cos(2\pi \bar{h}_A)\right)}{(h_A-1)^2(\bar h_A-1)^2}(1+f(h_A,\bar h_A))+W_0 \delta_{ij}\delta_{k\ell}~,
    \end{split}
    \label{gijklsumrule}
\end{equation}
where
\begin{equation}
    \begin{split}
        f(h,\bar h)=2(2-h-\bar h)+(h-\bar h)\left(\psi^{(0)}(\bar h)-\psi^{(0)}( h)\right)+(1-h)(1-\bar h)\left(\psi^{(1)}( h)+\psi^{(1)}( \bar h)\right)~,
    \end{split}
\end{equation}
where $\psi^{(m)}(z)$ is the polygamma function of order $m$. Again, for convergence of the $z,\bar z$ integrals near the origin, we are assuming no operators with $0\leq h_s<1$ or $0\leq\bar h_s< 1$ appear in the $s$-channel OPE. The constant $W_0$ gives the contribution from the vacuum Virasoro block when $i=j,k=\ell$. It vanishes at large $c$ since the double-discontinuity of the Virasoro vacuum block vanishes in this limit. Note that in this case, we must also subtract the contribution from the vacuum block from the sum, by expanding it in the $t$-channel OPE channel.

Let's consider the case $i=j\neq k=\ell$ of \eqref{gijklsumrule}, in which case we have
\begin{equation}
    \begin{split}
       R_{ikik}=-\frac{1}{8\pi^2}\sum_AC_{ik A}^2\frac{\left(1-\cos(2\pi \bar{h}_A)\right)}{(h_A-1)^2(\bar h_A-1)^2}(1+f(h_A,\bar h_A))+W_0~,
    \end{split}
    \label{Rikiksumrule}
\end{equation}
where we used \eqref{gijkl_4}. The function $f(h_A,\bar h_A)$ is negative for all operators but relevant scalar operators. Thus, if there are no relevant scalar operators in the $t$-channel, we find a simple bound for the sectional curvature $R_{ikik}$,
\begin{equation}
    \begin{split}
       R_{ikik}\geq -\frac{1}{8\pi^2}\sum_AC_{ik A}^2\frac{\left(1-\cos(2\pi \bar{h}_A)\right)}{(h_A-1)^2(\bar h_A-1)^2}+W_0~.
    \end{split}
    \label{gijklsumrule_bound1}
\end{equation}
We can do better if we impose stronger constraints on the $t$-channel OPE. The function $1+f(h_A,h_A)$ is negative for all $h_A\geq h_*\approx 1.30146 $. Hence, if all scalar operators in the $t$-channel OPE satisfy this, then we have 
\begin{equation}
    \begin{split}
       R_{ikik}\geq W_0~,
    \end{split}
    \label{gijklsumrule_bound2}
\end{equation}
which is a bound depending only on the central charge $c$. If the sum rules converge but there are scalar operators with $h_A\leq h_*$, then we can include the contribution from these operators explicitly, to find a lower bound to $R_{ikik}$. Finally, note that we can combine the sum rules \eqref{leadingsumrule_iikk} and \eqref{gijklsumrule_bound2}. To that end, assume that there are no relevant scalar operators in the $t$-channel OPE, and let $h=1+\epsilon$ be the holomorphic conformal weight of the lowest operator in the $t$-channel OPE. Adding $\mu$ times \eqref{leadingsumrule_iikk} to \eqref{Rikiksumrule}, we have 
\begin{equation}
    \begin{split}
       R_{ikik}=-\frac{1}{8\pi^2}\sum_AC_{ik A}^2\frac{\left(1-\cos(2\pi \bar{h}_A)\right)}{(h_A-1)^2(\bar h_A-1)^2}(1+f_\mu(h_A,\bar h_A))+(W_0+\mu V_0)~,
    \end{split}
    \label{Rikiksumrule_lambda}
\end{equation}
where
\begin{equation}
    \begin{split}
        f_\mu(h,\bar h)=2(\mu-1)(h+\bar h-2)+(h-\bar h)\left(\psi^{(0)}(\bar h)-\psi^{(0)}( h)\right)+(1-h)(1-\bar h)\left(\psi^{(1)}( h)+\psi^{(1)}( \bar h)\right)~.
    \end{split}
\end{equation}
We can choose $\mu=\mu_*(\epsilon)$ so that $1+f_{\mu_*(\epsilon)}(h,h)\leq0$ for all $h\geq 1+\epsilon$. Then we find the bound
\begin{equation}
    \begin{split}
       R_{ikik}\geq W_0+\mu_*(\epsilon) V_0~.
    \end{split}
    \label{Rikiksumrule_bound3}
\end{equation}
For example, when $\epsilon\ll1$, we have $\mu_*(\epsilon)=-\frac{1}{4\epsilon}+1+{\cal O}(\epsilon)$.

\subsubsection{$4d$}
\label{sec:4dgijklIF}

In $4d$, we find from \eqref{gijkl_EIF} and \eqref{LIF_1int} that the second derivative of the metric is given by
\begin{equation}
    \begin{split}
        g_{ij,k\ell}=\lim_{\delta\to0}\frac{1}{288\pi^2\delta}\int_0^1 dz d\bar z |z-\bar z|^2 G_{\Delta=3,J=\delta-3}(z,\bar z)\left(\left\langle[{\cal O}_k,{\cal O}_j][{\cal O}_i,{\cal O}_\ell]\right\rangle+(k\leftrightarrow\ell)\right).
    \end{split}
    \label{gijklsumrule1_4d}
\end{equation}
Using \eqref{4dCB}, some algebra gives
\begin{equation}
    \begin{split}
        G_{\Delta=3,J=\delta-3}(z,\bar z)&=\frac{3}{z-\bar z}\left[ \bar z (2-z)\ln (1-z)- z(2-\bar z)\ln(1-\bar z)\right]
        \\
        &+\frac{\delta}{2(z-\bar z)}\Big[\ln(1-z)\big(2(-8+7z)\bar z+3(-2+z)\bar z \ln z+3z \ln(1-\bar z)\big)
        \\
        &+6z \ln z\big(-3\bar z+(-2+\bar z)\ln(1-\bar z)\big)+6(-2+z) \bar z {\text{Li}}_2(z)-(z\leftrightarrow\bar z)\Big]~,
    \end{split}
    \label{CB4dgijkl}
\end{equation}
Let's consider the case $i=j\neq k=\ell$ for simplicity. The leading contribution of the conformal block \eqref{CB4dgijkl} contributes to the $1/\delta$ pole in \eqref{gijklsumrule1_4d}, which has to vanish by exact marginality. Thus, we find the naive sum rule:
\begin{equation}
    \begin{split}
        -\frac{32}{3\pi^2}\sum_{A}&C_{ik A}^2\left[1-\cos\left(\pi(\Delta_A+J_A)\right)\right](\Delta_A-4)
        \\
        &\times\frac{7J_A^4-2(44+3(\Delta_A-8)\Delta_A)J_A^2-(\Delta_A-6)^2(\Delta_A-2)^2}{\left[(\Delta_A-2)^2-J_A^2\right]^2\left[(\Delta_A-4)^2-J_A^2\right]^2\left[(\Delta_A-6)^2-J_A^2\right]^2}=0~,
    \end{split}
    \label{4dsumrule_leading}
\end{equation}
where we used the $t$-channel OPE \eqref{dDisc4d_tch}. As in $2d$, we have to be careful about the operators we are summing over. This is because the OPE expansion may not commute with the $z,\bar z$ integrals in \eqref{gijklsumrule1_4d}. The issue has to do with the behaviour of the double-discontinuity for small $z,\bar z$. At small $z,\bar z$, the kernel in \eqref{CB4dgijkl} goes to zero as $z$ (for fixed $\bar z$). Finiteness of the integrals at small $z$ is guaranteed provided that the integrals in the $s$-channel OPE converge, which requires $\Delta>J+4$ for all operators in the $s$-channel OPE. Note that this is above the unitarity bound $\Delta\geq J+2$.

Thus, for the sum rule \eqref{4dsumrule_leading} to be valid as written, we assume that the $s$-channel OPE expansion contains no operators that violate this bound. If such operators are present, we must include their contribution explicitly to the sum rule (which will give some additional contribution to \eqref{CB4dgijkl}), and also subtract their contribution to the $t$-channel sum in \eqref{CB4dgijkl}.

Assuming no $s$-channel operators with dimension $\Delta\leq J+4$, the sum rule \eqref{4dsumrule_leading} imposes constraints in the OPE of exactly marginal operators. Note that we do not have to include the contribution from the vacuum or stress tensor (global) blocks, because their double-discontinuity vanishes.

Let's now look at the finite piece in the inversion formula for $g_{ij,k\ell}$, coming from the subleading term in the conformal block expansion \eqref{CB4dgijkl}. Using the relation \eqref{gijkl_4}, we find
\begin{equation}
    \begin{split}
        R_{ikik}=-\frac{1}{96\pi^2}\sum_{A}C_{ik A}^2\left[1-\cos\left(\pi(\Delta_A+J_A)\right)\right] g(\Delta_A,J_A)~.
    \end{split}
    \label{4dsumrule_gijkl}
\end{equation}
The condition for convergence of this sum rule is the same as \eqref{4dsumrule_leading}, which we assume. Here, $g(\Delta,J)$ is the function \eqref{gfn4d}. Properties of this function are discussed in Appendix \ref{sec:4d_nasty}. If no operators with $\Delta<\Delta_*(J)$ appear in the $t$-channel OPE, where $\Delta_*(J)$ are given by \eqref{4dbadops}, then the sectional curvature satisfies
\begin{equation}
    \begin{split}
        R_{ikik}\geq 0~.
    \end{split}
    \label{Rijijbound1}
\end{equation}
Note that this condition is satisfied if all operators with $J=1$ or $J\geq3$ satisfy $\Delta\geq J+4$, and furthermore $J=0$ operators satisfy $\Delta\geq\Delta_*(J=0)\approx 4.43826$, and $J=2$ operators satisfy $\Delta\geq\Delta_*(J=2)\approx 6.05949$. As in $2d$, operators outside this range can be included and give a negative contribution to the RHS of \eqref{Rijijbound1}, provided the sum rule still converges.  Conversely, in regions of the conformal manifold with negative sectional curvature, operators with $\Delta <\Delta_*(J)$ necessarily appear in the $t$-channel OPE.

\section{Examples in $2d$}
\label{sec:examples}

In this section, we will verify some of the results discussed in section \ref{sec:inversion} in two examples in $2d$: free theories, and $(2,2)$ SCFTs. 

\subsection{$U(1)$ currents}
\label{sec:U1s}

Consider a $2d$ CFT with a holomorphic $U(1)$ current $j(z)$, and $N$ anti-holomorphic $U(1)$ currents $\tilde j_i(\bar z)$, $i=1,...,N$, with OPE
\begin{equation}
       j(z)j(0)\sim\frac{1}{z^2}~, \hspace{.2in}  \tilde j_i(\bar z)\tilde j_j(0)\sim \frac{\delta_{ij}}{\bar z^2}~.
\end{equation}
The exactly marginal operators are ${\cal O}_i(z,\bar z)\equiv j(z) \tilde j_i(\bar z)$, $i=1,...,N$. The Zamolodchikov metric $g_{ij}$ is given by
\begin{equation}
    \begin{split}
      g_{ij}\equiv\left\langle {\cal O}_i(e){\cal O}_j(0)\right\rangle=\delta_{ij}~.
        \end{split}
        \label{U1metric}
\end{equation}
An example of such a CFT is $N$ compact free bosons. In this case, the conformal manifold is $N^2$ dimensional, with $N$ choices for left/right-moving currents separately. The $N$ exactly marginal operators above define a $N$-dimensional conformal submanifold. 

Let's calculate the curvature of the tangent bundle. The 4-point function of exactly marginal operators is given by
\begin{equation}
    \begin{split}
      \left\langle {\cal O}_i(0){\cal O}_j(z,\bar z){\cal O}_k(1){\cal O}_\ell(\infty)\right\rangle=\left(\frac{1}{z^2}+\frac{1}{(1-z)^2}+1\right)\left(\frac{\delta_{ij}\delta_{k\ell}}{\bar z^2}+\frac{\delta_{i\ell}\delta_{jk}}{(1-\bar z)^2}+\delta_{ik}\delta_{j\ell}\right)~,
        \end{split}
\end{equation}
and the connected part is
\begin{equation}
    \begin{split}
      \left\langle {\cal O}_i(0){\cal O}_j(z,\bar z){\cal O}_k(1){\cal O}_\ell(\infty)\right\rangle_c&=\frac{1}{z^2}\left(\frac{\delta_{i\ell}\delta_{jk}}{(1-\bar z)^2}+\delta_{ik}\delta_{j\ell}\right)
      \\
      &+\frac{1}{(1-z)^2}\left(\frac{\delta_{ij}\delta_{k\ell}}{\bar z^2}+\delta_{ik}\delta_{j\ell}\right)+\left(\frac{\delta_{ij}\delta_{k\ell}}{\bar z^2}+\frac{\delta_{i\ell}\delta_{jk}}{(1-\bar z)^2}\right)~,
        \end{split}
        \label{U14ptc}
\end{equation}

Using \eqref{U14ptc} in the Euclidean inversion formula, we find
\begin{equation}
       R_{ijk\ell}=\frac{1}{4}\left(\delta_{i\ell}\delta_{jk}-\delta_{ik}\delta_{j\ell}\right)=\frac{1}{4}\left(g_{i\ell}g_{jk}-g_{ik}g_{j\ell}\right)~,
        \label{U1curvature}
\end{equation}
as expected for a homogeneous manifold of constant negative curvature. To obtain \eqref{U1curvature}, we used the integrals
\begin{equation}
\int d^2 z \ln |1-z|^2 \frac{1}{z^2}=-\pi~, \hspace{.1in}\int d^2 z\ln |1-z|^2 \frac{1}{z^2 (1-\bar z)^2}=\pi~, \hspace{.1in}\int d^2 z \ln |1-z|^2 \frac{1}{(1-z)^2}=0~.
  \label{zintEIF}
\end{equation}

Next we turn to the Lorentzian inversion formula. The double-discontinuity of \eqref{U14ptc} is subtle to compute, because naively it vanishes since the the expansion in powers of $(1-\bar z)$ only has integer powers. However, it turns out that the correct answer is that the double-discontinuity of \eqref{U14ptc} has delta-function localized contributions \cite{Caron-Huot:2017vep}. To see this, let's first compute the double-discontinuity of $z^{-2}(1-\bar z)^{-2}$, starting from
\begin{equation}
    \begin{split}
        \mathrm{dDisc}\left[\frac{1}{z^2\bar{z}^2}\left(\frac{1-\bar{z}}{\bar z}\right)^a\right]=(1-\cos\left(2\pi a\right))\frac{1}{z^2\bar{z}^2}\left(\frac{1-\bar{z}}{\bar z}\right)^a~.
    \end{split}
    \label{ddisc1}
\end{equation}
By taking $a\to-2$, we get the double discontinuity of $z^{-2}(1-\bar z)^{-2}$.

From \eqref{LIF_1int} and \eqref{CB2d},  we are interested in the kernel
\begin{equation}
    \begin{split}
        G_{J+1,\Delta-1}(z,\bar z)=k_{2h}(z)k_{2-2\bar h}(\bar z)+k_{2-2\bar h}(z)k_{2 h}(\bar z)~,
    \end{split}
    \label{2dGLIF}
\end{equation}
where $h=\frac{\Delta+J}{2}, \bar h=\frac{\Delta-J}{2}$. Using the integral representation of ${}_2F_1$, we can show that \cite{Caron-Huot:2017vep}
\begin{equation}
    \begin{split}
        \int_0^1 d\bar z~k_{2-2\bar h}(\bar z){\mathrm{dDisc}}\left[\frac{1}{\bar{z}^2}\left(\frac{1-\bar{z}}{\bar z}\right)^a\right]=-\pi\frac{1-\cos(2\pi a)}{\sin(\pi(a+\bar h))}\frac{\Gamma(1+a)^2\Gamma(2-2\bar h)}{\Gamma(1-\bar h)^2\Gamma(2+a-\bar h)\Gamma(1+a+\bar h)}~.
        \end{split}
        \label{zbardDisc}
\end{equation}
As advertised, the integral is finite at $a\to-2$. The $z$ integral gives
\begin{equation}
    \begin{split}
        \int_0^1 d z~k_{2 h}( z)\frac{1}{z^2}=\frac{2^{-1+2h}}{(h-1)\sqrt{\pi}}\frac{\Gamma\left(\frac{1}{2}+h\right)}{\Gamma\left(1+h\right)}~.
        \end{split}
\end{equation}
Using these two results, we find
\begin{equation}
    \begin{split}
        \int_0^1 dz \int_0^1 d\bar z~k_{2 h}( z)k_{2-2\bar h}(\bar z)&{\mathrm{dDisc}}\left[\frac{1}{z^2\bar{z}^2}\left(\frac{1-\bar{z}}{\bar z}\right)^{-2}\right]
        \\
        &~~=2\pi^2\bar h(\bar h-1)\frac{\Gamma(2-2\bar h)}{\Gamma(1-\bar h)^2}\frac{2^{-1+2h}}{(h-1)\sqrt{\pi}}\frac{\Gamma\left(\frac{1}{2}+h\right)}{\Gamma\left(1+h\right)}~.
        \end{split}
        \label{zzbardDisc}
\end{equation}
Finally, we set $J=1$ followed by $\Delta=1$ to find\footnote{From \eqref{zzbardDisc}, we see that for fixed $J$ close to 1, there is a pole in $\Delta$ at $2-J$, with residue proportional to $J-1$. If we take $\Delta\to1$ before $J\to1$, the pole and finite part mix and give the wrong answer.}
\begin{equation}
    \begin{split}
        \left.\int_0^1 dz \int_0^1 d\bar z~k_{2 h}( z)k_{2-2\bar h}(\bar z){\mathrm{dDisc}}\left[\frac{1}{z^2\bar{z}^2}\left(\frac{1-\bar{z}}{\bar z}\right)^{-2}\right]\right|_{J=1,\Delta=1}=-2\pi^2~.
        \end{split}
        \label{2FBzzbarint}
\end{equation}

We can proceed similarly to find the double-discontinuities of the other terms in \eqref{U14ptc}. We find that, at $\Delta=1,J=1$, all these double-discontinuities vanish except for $z^{-2}(1-\bar z)^{-2}$ and $\bar z^{-2}(1- z)^{-2}$, which give the same contribution. Thus, using these results in \eqref{LIF_1int}, we get
\begin{equation}
    \begin{split}
        R_{ijk\ell}&=-\frac{1}{32\pi^2}\left[-(8\pi^2)\left(\delta_{i\ell}\delta_{jk}+\delta_{ij}\delta_{k\ell}\right)+(8\pi^2)\left(\delta_{ik}\delta_{j\ell}+\delta_{ij}\delta_{k\ell}\right)\right]
        \\
        &=\frac{1}{4}\left(\delta_{i\ell}\delta_{jk}-\delta_{ik}\delta_{j\ell}\right)~,
    \end{split}
    \label{RU1}
\end{equation}
where the first minus sign comes from $\alpha_{\Delta=1,J=1}$ in \eqref{LIF_1int}, and the two terms in the square brackets come from the two contributions in \eqref{LIF_1int}. There is also an overall factor of $(-2)$ relative to \eqref{2FBzzbarint} coming from \eqref{dDisc1}, and another factor of 2 from the two terms in \eqref{2dGLIF}. In the end, we find perfect agreement with \eqref{U1curvature}.

Let's now check the inversion formula for $g_{ij,k\ell}$ \eqref{gijkl_EIF} using the Lorentzian inversion formula. Doing a similar analysis as the one leading to \eqref{2FBzzbarint}, we find
\begin{equation}
    \begin{split}
       &\left.\int_0^1 dz \int_0^1 d\bar z~G_{J+1,\Delta-1}(z,\bar z){\mathrm{dDisc}}\left[\frac{1}{z^2\bar{z}^2}\left(\frac{1-\bar{z}}{\bar z}\right)^{-2}\right]\right|_{J=0}=\pi^2\Delta+{\cal O}(\Delta^2)~,
       \\
       &\left.\int_0^1 dz \int_0^1 d\bar z~G_{J+1,\Delta-1}(z,\bar z){\mathrm{dDisc}}\left[\frac{1}{z^2\bar{z}^2}\left(\frac{1-z}{z}\right)^{-2}\right]\right|_{J=0}=\pi^2\Delta+{\cal O}(\Delta^2)~,
       \\
       &\left.\int_0^1 dz \int_0^1 d\bar z~G_{J+1,\Delta-1}(z,\bar z){\mathrm{dDisc}}\left[\frac{1}{(1-z)^2}\right]\right|_{J=0}=-\pi^2\Delta+{\cal O}(\Delta^2)~,
       \\
       &\left.\int_0^1 dz \int_0^1 d\bar z~G_{J+1,\Delta-1}(z,\bar z){\mathrm{dDisc}}\left[\frac{1}{(1-\bar z)^2}\right]\right|_{J=0}=-\pi^2\Delta+{\cal O}(\Delta^2)~,
        \end{split}
        \label{ddiscsgijkl}
\end{equation}
while the other terms in \eqref{U14ptc} have zero double-discontinuity. The prefactor $\alpha_{\Delta,J}$ is given by
\begin{equation}
    \begin{split}
        \alpha_{\Delta,J=0}=-\frac{1}{2\Delta}\frac{\hat C_J(1)}{{\text{vol}}SO(d-1)}+{\cal O}(\Delta^0)~.
    \end{split}
    \label{prefacgijkl}
\end{equation}
Using \eqref{U14ptc}, \eqref{ddiscsgijkl}, and \eqref{prefacgijkl} in \eqref{LIF_1int} and \eqref{gijkl_EIF}, we find
\begin{equation}
    \begin{split}
        g_{ij,k\ell}&=\lim_{\Delta\to0}\frac{1}{24\pi^2}\frac{2}{\Delta}\pi^2\Delta\left[\left(\delta_{ij}\delta_{k\ell}-\delta_{ik}\delta_{j\ell}\right)+\left(\delta_{ij}\delta_{k\ell}-\delta_{i\ell}\delta_{jk}\right)\right]
        \\
        &=\frac{1}{12}\left(2\delta_{ij}\delta_{k\ell}-\delta_{i\ell}\delta_{jk}-\delta_{ik}\delta_{j\ell}\right)~.
    \end{split}
\end{equation}
This is the expression for the second derivative of the metric in Riemann normal coordinates, as can be seen from \eqref{gijkl_4} and \eqref{RU1}. Note also that the double-discontinuities \eqref{ddiscsgijkl} vanish as $\Delta\to0$, as required for exactly marginal operators, see \eqref{gijkl_intgd}.

Finally, let's consider the sum rule \eqref{zam2dsumrule_all}. Since this sum rule does not converge if there are holomorphic conserved currents in the $s$-channel, we have to subtract their contribution from the correlation function \eqref{U14ptc}. The remaining contribution to the 4-point function has vanishing double-discontinuity. Meanwhile, the contribution of the $s$-channel holomorphic currents has to be included explicitly, in the same way as the analysis leading up to \eqref{RU1}. Thus, in the end, we obtain again the result \eqref{RU1}.

A similar analysis holds for the sum rule for $g_{ij,k\ell}$.

\subsection{$2d$ $(2,2)$ SCFTs}

Now we will compute the curvature of various vector bundles over the conformal manifold of $2d$ $(2,2)$ SCFTs. We will summarize the basic facts that we need about such SCFTs here, for a more extended discussion see e.g. \cite{Lerche:1989uy}. The symmetry algebra of such a CFT is enhanced to ${\cal N}=2$ super-Virasoro algebra, with left-moving supercurrents $G^\pm(z)$ and left-moving $R$-current $J(z)$, as well as right-moving analogues.

The (super-)primaries with respect to the super-Virasoro algebra are annihilated by all lowering operators. Furthermore, multiplets can be long or short. In short multiplets, primaries are further annihilated by either $G^+_{-1/2}$ (chiral) or $G^-_{-1/2}$ (anti-chiral) in the left-moving part, and similarly in the right-moving part. This gives four rings of chiral primaries: $(c,c)$, $(a,c)$, $(c,a)$, and $(a,a)$, where the notation means chiral/anti-chiral in the left- or right-moving part. Operators in the $(c,c)$ chiral ring satisfy $J_0=2L_0$, $\bar J_0=\bar 2L_0$, in the $(a,c)$ ring they satisfy $J_0=-2L_0$, $\bar J_0=2\bar L_0$, etc. The $(c,c)$ operators together with their complex conjugate form the chiral ring, while the $(a,c)+(c,a)$ operators comprise the twisted chiral ring. 

The chiral operators with $J_0=\pm1$ and $\bar J_0=\pm1$ are especially important for us. Acting with $G^-_{-\frac{1}{2}}\bar G^-_{-\frac{1}{2}}$ on the $(c,c)$ chiral primary with $J_0=\bar J_0=1$ gives an exactly marginal operator with $R$-charges $J_0=\bar J _0=0$. Similarly, there are exactly marginal operators obtained by acting with $G^+_{-\frac{1}{2}}\bar G^+_{-\frac{1}{2}}$ on $(a,a)$ chiral operators. Altogether, these form the chiral moduli space. There is also the twisted chiral moduli space, where the exactly marginal operators are obtained by acting with supercurrents on operators in the twisted chiral ring.

The moduli space of the SCFT locally factorizes in the chiral moduli space, and the twisted chiral moduli space. The geometry of the conformal manifold is known to be that of a complex K\"ahler manifold. (See e.g.\ \cite{Seiberg:1988pf})

The dimension of chiral operators is constant over the conformal manifold, and hence they form a vector bundle. Furthermore, chiral operators can only mix with other chiral operators of the same $R$-charge. The chiral ring coefficients define a multiplication between different chiral bundles which corresponds to taking the OPE between the operators. These chiral ring coefficients depend holomorphically on the coordinates on the conformal manifold.

From now on, we will consider only the chiral moduli space for simplicity, though all computations have analogues in the twisted moduli space. The two-point function between a $(c,c)$ chiral primary operator $\varphi_i$ of $R$-charge $q$ and a $(a,a)$ chiral primary operator $\varphi_{\bar j}$ of $R$-charge $-q$ is given by
\begin{equation}
    \begin{split}
        h^{(q)}_{i\bar j}(0)\equiv \left\langle\varphi_i(1)\varphi_{\bar j}(0)\right\rangle~.
    \end{split}
\end{equation}
We use conventions where unbarred indices denote chiral operators in the $(c,c)$ ring, and barred coordinates give chiral operators in the $(a,a)$ ring. The two-point function is non-zero only between $(c,c)$ and $(a,a)$ operators. The OPE between chiral operators $\varphi_i$ is given by
\begin{equation}
    \begin{split}
        &\varphi_i(z)\varphi_j(0)=C_{ij}^m \varphi_m(0)+(\text{subleading)}~,
        \end{split}
        \label{ccOPE}
        \end{equation}
and similarly for anti-chiral operators. We will also need the OPE between a $(c,c)$ chiral operator of $R$-charge $q_i$, $\varphi_i$, and a $(a,a)$ operator of $R$-charge $-q_{j}$, $\varphi_{\bar j}$. When $q_i\leq q_j$, the OPE is given by
\begin{equation}
    \begin{split}
        \varphi_i(z)\varphi_{\bar j}(0)\sim &|z|^{-2q_i} C_{i k}^{p}h^{(q_j)}_{p\bar j}h^{(q_k)k\bar m}\varphi_{\bar m}(0)
        \\
        &~~~~~~+\frac{3 q_i}{c}h^{(q_i)}_{i\bar j}\delta_{q_i,q_j}\left(z^{-\Delta_\varphi+1}\bar z^{-\Delta_\varphi}J(0)+\bar z^{-\Delta_\varphi}\bar z^{-\Delta_\varphi+1}\bar J(0)\right)+(\cdots)~,
    \end{split}
    \label{acOPE}
\end{equation}
where $\varphi_{\bar m}$ is a $(a,a)$ operator of $R$-charge $-(q_j-q_i)$. We assume the $(c,c), (a,a)$ operators have the same left and right $R$-charges, for simplicity. We included the contributions from the anti-chiral operators together with the $R$-currents $J(z)$ and $\bar J(\bar z)$, which will be important for us. When $q_i\geq q_j$, this OPE becomes
\begin{equation}
    \begin{split}
        \varphi_i(z)\varphi_{\bar j}(0)\sim &|z|^{-2q_j} C_{\bar j \bar \ell }^{\bar p}h^{(q_i)}_{i\bar p}h^{(q_\ell) m\bar n}\varphi_{ m}(0)
        \\
        &~~~~~~+\frac{3 q_i}{c}h^{(q_i)}_{i\bar j}\delta_{q_i,q_j}\left(z^{-\Delta_\varphi+1}\bar z^{-\Delta_\varphi}J(0)+\bar z^{-\Delta_\varphi}\bar z^{-\Delta_\varphi+1}\bar J(0)\right)+(\cdots)~.
    \end{split}
    \label{acOPE2}
\end{equation}
For more details, see e.g. \cite{deBoer:2008ss}.

The Zamolodchikov metric is given by
\begin{equation}
    \begin{split}
        g_{i\bar j}(0)\equiv\langle \left(G^+_{-\frac{1}{2}}\bar G^+_{-\frac{1}{2}} \phi_{\bar j}\right)(1)\left(G^-_{-\frac{1}{2}}\bar G^-_{-\frac{1}{2}} \phi_{i}\right)(0)\rangle=4h^{(1)}_{i\bar j}(0)~,
    \end{split}
    \label{gijN2chiral}
\end{equation}
where $\phi_i$ denote the chiral primaries of $R$-charge $J_0=1$. The $i,\bar i$ indices give K\"ahler coordinates on the conformal manifold.

\subsubsection{Curvature of Chiral Primaries}

We first compute the curvature of the chiral vector bundle using the Euclidean inversion formula. From \eqref{Rijkl_EIF1int_2d}, the correlation function of interest is
\begin{equation}
    \begin{split}
        \left\langle\varphi_k(0)\varphi_{\bar\ell}(z,\bar z){\cal O}_{i}(1){\cal O}_{\bar j}(\infty)\right\rangle~.
    \end{split}
\end{equation}
Using superconformal Ward identities, we have
\begin{equation}
    \begin{split}
        \left\langle\varphi_k(0)\varphi_{\bar\ell}(z,\bar z){\cal O}_{i}(1){\cal O}_{\bar j}(\infty)\right\rangle&=|z|^{-2\Delta_\varphi}\left\langle  {\cal O}_{\bar j}(0)\varphi_{\bar \ell}(1/z,1/\bar z){\cal O}_i(1){\varphi}_{k}(\infty)\right\rangle~,
        \\
        &=|z|^{4-2\Delta_\varphi}\left\langle  {\cal O}_{\bar j}(0){\cal O}_i(z,\bar z)\varphi_{\bar \ell}(1){\varphi}_{k}(\infty)\right\rangle~,
        \\
        &=4|z|^{4-2\Delta_\varphi}\left\langle  (\partial\bar\partial{ \phi}_{\bar j})(0){ \phi}_i(z,\bar z)\varphi_{\bar \ell}(1){\varphi}_{k}(\infty)\right\rangle~,
    \end{split}
\end{equation}
where in the last line we wrote ${\cal O}_i=G^-_{-1/2}\phi_i$ as a contour of the supercurrent $G^-(w)$ around $w=z$, and deformed the contour. Since $\varphi_{\bar\ell}$ is anti-chiral, it is annihilated by $G^-_{-1/2}$. Similarly, $G^-(w)$ annihilates $\varphi_k(\infty)$ since the operator is at infinity. The only contribution comes from $G^-(w)$ acting on ${\cal O}_{\bar j}$, which gives $G^-_{-\frac{1}{2}}G^+_{-\frac{1}{2}}\phi_{\bar j}=2L_{-1}\phi_{\bar j}$. A similar argument holds for the anti-holomorphic part. Using another conformal transformation, we have
\begin{equation}
    \begin{split}
       \left\langle  (\partial\bar\partial{ \phi}_{\bar j})(0){ \phi}_i(z,\bar z)\varphi_{\bar \ell}(1){\varphi}_{k}(\infty)\right\rangle=\partial\bar\partial\left[|1-z|^2  \left\langle  { \phi}_{\bar j}(0){ \phi}_i(z,\bar z)\varphi_{\bar \ell}(1){\varphi}_{k}(\infty)\right\rangle\right]~.
    \end{split}
\end{equation}
Using this in \eqref{Rijkl_EIF1int_2d}, we have
\begin{equation}
    \begin{split}
         F_{i\bar j}{}^{\bar m}{}_{\bar \ell}(0)&=\frac{h^{(q_k)~\bar m k}}{4\pi}\int d^2 z\ln|1-z|^2 |z|^{-4+2\Delta_{\varphi}}\left\langle\varphi_k(0)\varphi_{\bar\ell}(z,\bar z){\cal O}_{i}(1){\cal O}_{\bar j}(\infty)\right\rangle_c~,
         \\
         &=\frac{h^{(q_k)~\bar m k}}{\pi}\int d^2 z\ln|1-z|^2\partial\bar\partial\left[|1-z|^2  \left\langle  { \phi}_{\bar j}(0){ \phi}_i(z,\bar z)\varphi_{\bar \ell}(1){\varphi}_{k}(\infty)\right\rangle\right]-F_{\text{disc}}~.
    \end{split}
    \label{Rijkl_N2_2d_b}
\end{equation}
Here, $F_{\text{disc}}$ stands for the contribution to $F_{i\bar j}{}^{\bar m}{}_{\bar \ell}(0)$ coming form the $s$-channel disconnected contribution to the 4-point function in the first line, namely
\begin{equation}
    \begin{split}
        F_{\text{disc}}=\frac{h^{(q_k)~\bar m k}}{\pi}\int d^2 z\ln|1-z|^2 |z|^{-4}h^{(q_k)}_{k\bar\ell}h^{(1)}_{i\bar j}~.
    \end{split}
    \label{Fdisc}
\end{equation}
The $t,u$-channel disconnected contributions vanish. Using $\nabla^2=4\partial\bar\partial$, \eqref{Rijkl_N2_2d_b} can be written as
\begin{equation}
    \begin{split}
              F_{i\bar j}{}^{\bar m}{}_{\bar \ell}(0)&= \frac{h^{(q_k)~\bar m k}}{4\pi}\int d^2 z\nabla\cdot\left(\ln|1-z|^2\nabla\left[|1-z|^2  \left\langle  { \phi}_{\bar j}(0){ \phi}_i(z,\bar z)\varphi_{\bar \ell}(1){\varphi}_{k}(\infty)\right\rangle\right]\right)
               \\
               &-\frac{h^{(q_k)~\bar m k}}{4\pi}\int d^2 z\nabla\cdot\left(\left[\nabla\ln|1-z|^2\right]|1-z|^2  \left\langle  { \phi}_{\bar j}(0){ \phi}_i(z,\bar z)\varphi_{\bar \ell}(1){\varphi}_{k}(\infty)\right\rangle\right)-F_{\text{disc}}~,
    \end{split}
    \label{Rijkl_N2_2d_c}
\end{equation}
where we used the fact that $\partial\bar\partial\ln|1-z|^2=0$. The non-vanishing part of the first two terms in \eqref{Rijkl_N2_2d_c} comes only from the boundary terms at $z=0,1,\infty$. 

Let's ignore for now the disconnected contribution in the correlation function in \eqref{Rijkl_N2_2d_c}, and look at the second term in \eqref{Rijkl_N2_2d_c}. Consider the contribution from the boundary term $|z|=L$ at infinity. Using conformal transformations, we have
\begin{equation}
    \begin{split}
        |1-z|^2  \left\langle  { \phi}_{\bar j}(0){ \phi}_i(z,\bar z)\varphi_{\bar \ell}(1){\varphi}_{k}(\infty)\right\rangle&=\frac{|1-z|^2}{|z|^2}\left\langle  {\varphi}_{k}(0){ \phi}_i(1/z,1/\bar z)\varphi_{\bar \ell}(1){ \phi}_{\bar j}(\infty)\right\rangle
        \\
        &=C_{ik}^m C_{\bar j\bar \ell}^{\bar n}h^{(q_k+1)}_{m\bar n}+({\text{subleading}})~,
    \end{split}
\end{equation}
up to terms that are subleading at large $|z|$. Letting $z=r e^{i\theta}$, we also have
\begin{equation}
    \begin{split}
        \nabla\left[\ln|1-z|^2\right]=\hat r\left(\frac{2}{r}+{\text{subleading}}\right)+\hat\theta\left(\cdots\right)~,
    \end{split}
\end{equation}
where we dropped terms subleading at large $r$ and the $\cdots$ on the RHS denote terms that will not be important for us. Stoke's theorem gives the contribution from the region $|z|=L\gg1$ to be
\begin{equation}
    \begin{split}
        -\frac{h^{(q_k)~\bar m k}}{4\pi} \int_0^{2\pi} d\theta ~L \hat r\cdot&\left(\frac{2}{L} C_{ik}^m C_{\bar j\bar \ell}^{\bar n}h^{(q_k+1)}_{m\bar n}\hat r+\hat r({\text{subleading}})+\hat\theta(\cdots)\right)
        \\
        &~~~~~~~~~~~~~~~~~~~~~~~~~~~~~~~~~~=-C_{ik}^m C_{\bar j\bar \ell}^{\bar n}h^{(q_k)~\bar m k}h^{(q_k+1)}_{m\bar n}~,
    \end{split}
    \label{2dN2bdry_zinfty}
\end{equation}
where in the last line we used the OPE \eqref{ccOPE} and took the limit $L\to\infty$.

The contribution from the boundary term near $z=1$ to the second term in \eqref{Rijkl_N2_2d_c} can be obtained similarly. Letting $1-z=r e^{i\theta}$, Stoke's theorem gives
\begin{equation}
    \begin{split}
        -\frac{h^{(q_k)~\bar m k}}{4\pi} \int_0^{2\pi} d\theta ~\epsilon (-\hat r)\cdot&\left(\frac{2}{\epsilon} \hat r\right)\left.\left(|1-z|^2  \left\langle  { \phi}_{\bar j}(0){ \phi}_i(z,\bar z)\varphi_{\bar \ell}(1){\varphi}_{k}(\infty)\right\rangle\right)\right|_{z=1}
        \\
        &~~~~~~~~~~~~~~~~~~~~~~=C_{i n}^{p}C_{\bar j\bar s}^{\bar m}h^{(q_p)}_{p\bar \ell}h^{(q_n)n\bar s}~,
    \end{split}
    \label{2dN2bdry_z1}
\end{equation}
where we used the OPE \eqref{acOPE} to evaluate the correlation function. Note that the contribution from \eqref{acOPE2} vanishes at $z=1$. 

Finally let's evaluate the contribution from the boundary term at $z=0$. Let $z=r e^{i\theta}$, so that
\begin{equation}
\begin{split}
    \nabla \ln|1-z|^2=\hat r\left(-2\cos\theta-2r\cos(2\theta)\right)+\hat\theta\left(2r\sin\theta\right)+\cdots~.
\end{split}    
\end{equation}
Using this together with the OPE \eqref{acOPE}, we find the boundary contribution
\begin{equation}
    \begin{split}
        -\frac{h^{(q_k)~\bar m k}}{4\pi} \int_0^{2\pi} d\theta ~\epsilon (-\hat r)\cdot&\left(-2\hat r\cos\theta-2\hat r\epsilon\cos(2\theta)\right)\left(1+\epsilon^2-2\epsilon\cos\theta\right) h^{(1)}_{i\bar j}h^{(q_k)}_{k\bar \ell}\left(\frac{1}{\epsilon^2}+\frac{3 q_k}{c} \frac{2\cos\theta}{\epsilon}\right)
        \\
        &=h^{(1)}_{i\bar j}\delta^{\bar m}_{\bar\ell}\left(1-\frac{3 q_k}{c}\right)~.
    \end{split}
    \label{2dN2bdry_z0}
\end{equation}
One can check that other contributions to the OPE, no written explicitly in \eqref{ccOPE} and \eqref{acOPE}, do not contribute to the integral in \eqref{2dN2bdry_z0}.

Now let's discuss the first term in \eqref{Rijkl_N2_2d_c}, ignoring the disconnected contribution to the correlation function for a moment. Once again, the integral localizes at $z=0,1,\infty$. However, now we see that the derivative acts on the correlation function. Proceeding as above, we see that this kills the boundary contributions at $z=1,\infty$. On the other hand, the boundary contribution at $z=0$ is once again given by \eqref{2dN2bdry_z0}.

Finally, there is the disconnected contribution \eqref{Fdisc}. Cutting out the integral near $|z|>\epsilon$, the integral can be carried out explicitly to give
\begin{equation}
    \begin{split}
        F_{\text{disc}}=h^{(1)}_{i\bar j}\delta^{\bar m}_{\bar \ell}~.
    \end{split}
\end{equation}
Combining this with the boundary contributions at $z=0,1,\infty$ to \eqref{Rijkl_N2_2d_c}, we find
\begin{equation}
    \begin{split}
         F_{i\bar j}{}^{\bar m}{}_{\bar \ell}(0)=-C_{ik}^m C_{\bar j\bar \ell}^{\bar n}h^{(q_k)~\bar m k}h^{(q_k+1)}_{m\bar n}+C_{i n}^{p}C_{\bar j\bar s}^{\bar m}h^{(q_j)}_{p\bar \ell}h^{(q_n)n\bar s}+h^{(1)}_{i\bar j}\delta^{\bar m}_{\bar\ell}\left(1-\frac{6 q_k}{c}\right)~.
    \end{split}
    \label{Fchiral}
\end{equation}

\subsubsection{Curvature of Supercurrents}

As we deform the CFT along the conformal manifold, the supercurrents $G^\pm(z)$ can pick up a phase $G^\pm(z)e^{i\theta(\lambda)}$ (and similarly for the right-moving supercurrents). This phase only acts on the supercurrents, and therefore leaves the $N=2$ superconformal algebra invariant.

We can think of the supercurrents as defining a line bundle over the conformal manifold. The connection on this line-bundle dictates how the operators transform as we move on the conformal manifold. For the supercurrent, we have
\begin{equation}
    \begin{split}
      {\cal O}_i(x)  G^+(0)\sim A_i(\lambda)\delta^{(2)}(x)G^+(0)~,
    \end{split}
\end{equation}
where $A_i(\lambda)$ is the connection for the supercurrent $G^+$. Since we want to leave the two-point function
\begin{equation}
    \begin{split}
        \langle G^-(1)G^+(0)\rangle_\lambda~,
    \end{split}
\end{equation}
unchanged, it follows that 
\begin{equation}
    \begin{split}
      {\cal O}_i(x)  G^-(0)\sim -A_i(\lambda)\delta^{(2)}(x)G^-(0)~.
    \end{split}
\end{equation}
Using \eqref{Rijkl_EIF1int_2d}, the curvature of the supercurrent vector bundle is given by\footnote{Since the supercurrents are not scalar operators, the integrand of \eqref{Rijkl_EIF1int_2d} has to be modified slightly, see \cite{Kravchuk:2018htv}. Alternatively, we can compute the curvature of the vector bundle of the scalars operators $G^\pm\bar G^\pm$, which is equal to \eqref{Fkl} times a factor of 2.}
\begin{equation}
    \begin{split}
        F^{G}_{k\bar\ell}=\frac{1}{4\pi}\frac{3}{2c}\int d^2 z\frac{z^3}{|z|^4}\ln|1-z|^2\langle G^-(0)G^+(z){\cal O}_k(1){\cal O}_{\bar \ell}(\infty)\rangle_c~,
    \end{split}
    \label{Fkl}
\end{equation}
where the factor of $\frac{3}{2 c}$ comes from dividing by the two-point function $\langle G^-(1) G^+(0)\rangle=\frac{2c}{3}$, as needed to raise one of the indices, see \eqref{Rijkl_EIF1int_2d}. 

The 4-point function $\langle G^-(0)G^+(z){\cal O}_k(1){\cal O}_{\bar \ell}(\infty)\rangle$ can be evaluated using superconformal ward-identities. The only operators appearing in $s$-channel OPE are the identity and the stress-tensor. Thus we have
\begin{equation}
    \begin{split}
        \langle G^-(0)G^+(z){\cal O}_k(1){\cal O}_{\bar \ell}(\infty)\rangle_c=-\frac{8}{z}h^{(1)}_{k\bar\ell}~.
    \end{split}
\end{equation}
Using this in \eqref{Fkl}, and using the integral \eqref{zintEIF}, we find
\begin{equation}
    \begin{split}
F^G_{k\bar\ell}=\frac{3h^{(1)}_{k\bar\ell}}{c}~.
    \end{split}
    \label{Fkl_final}
\end{equation}

\subsubsection{Riemann Curvature}

The exactly marginal operators are top components of the chiral/anti-chiral operators. To compute the Riemann curvature of the Zamolodchikov metric, we have to combine the curvature of chiral primaries of $R$-charge $J_0=\bar J_0=\pm1$, given by \eqref{Fchiral}, with the curvature of the supercurrents, \eqref{Fkl_final}. 

If the operators $\varphi_k$,$\varphi_{\bar\ell}$ have $R$-charges $J_0=\bar J_0=\pm1$, the contribution from the second term in \eqref{Fchiral} is equal to $h^{(1)}_{i\bar\ell}\delta^{\bar m}_{\bar j}$. 
On the other hand, there is a left-moving and a right-moving copy of the supercurrents, each with a curvature \eqref{Fkl_final}. There is also a factor of 4 relating the two-point function of the chiral primaries and that of exactly marginal operators, \eqref{gijN2chiral}. In total, we find the Riemann curvature
\begin{equation}
    \begin{split}
        R_{i\bar jk\bar\ell}&=-4C_{ik}^m C_{\bar j\bar \ell}^{\bar n}h^{(2)}_{m\bar n}+4h^{(1)}_{i\bar\ell}h^{(1)}_{k\bar j}+4h^{(1)}_{i\bar j}h^{(1)}_{k\bar \ell}.
        \\
        &=-4C_{ik}^m C_{\bar j\bar \ell}^{\bar n}h^{(2)}_{m\bar n}+\frac{1}{4}\left(g_{i\bar\ell}g_{k\bar j}+g_{i\bar j}g_{k\bar \ell}\right).
    \end{split}
    \label{R2dSCFT22}
\end{equation}

Note that the curvature satisfies
\begin{equation}
    \begin{split}
        R_{i\bar i k\bar k}&=-4C_{ik}^m C_{\bar i\bar k}^{\bar n}h_{m\bar n}+\frac{g_{i\bar k}g_{k\bar i}+g_{i\bar i}g_{k\bar k}}{4}
        \\
        &\leq\frac{g_{i\bar k}g_{k\bar i}+g_{i\bar i}g_{k\bar k}}{4}~.
    \end{split}
    \label{Rbound_N2}
\end{equation}
In particular, when the central charge $c$ is an integer multiple of 3, the bound  \eqref{Rbound_N2} is sharp and saturated by the free theory of $c/3$ complex bosons+fermions. This may be deduced from a parallel analysis to that of section \ref{sec:U1s}.

More generally, \eqref{Rbound_N2} is in qualitative agreement with there being a lower bound on the sectional curvature, as seen in the sum rule \eqref{gijklsumrule_bound2}.\footnote{We caution that Riemann normal coordinates are not the same as the K\"ahler coordinates used in \eqref{Rbound_N2}.}  For example, by contracting indices \eqref{Rbound_N2} yields a general lower bound on the scalar curvature of the Zamolodchikov metric
\begin{equation}\label{scalarboundN2}
    R\geq -\frac{n(n+1)}{4}~,
\end{equation}
where above, $n$ is the complex dimension of the conformal manifold.  By averaging these inequalities, we can further constrain the topology and geometry of the manifold.  For instance in the special case $n=1$, a smooth compact manifold has a genus $h$ computed by the Gauss-Bonnet theorem:
\begin{equation}
    \frac{1}{4\pi}\int \sqrt{g}R=2-2h~.
\end{equation}
From \eqref{scalarboundN2} we then find that the volume $\text{vol}$ of the conformal manifold satisfies:
\begin{equation}
    \mathrm{vol}\geq 16\pi (h-1)~.
\end{equation}
Such volumes of the space of conformal field theories have been previously investigated in  \cite{Moore:2015bba}.  Analogous results may be proven in the context of $4d$ $N=2$ theories using the results of \cite{Papadodimas:2009eu, Baggio:2014ioa}.

\section*{Acknowledgements}

We thank N. Afkhami-Jeddi, A. Ashmore, S. Chowdhury, D. Kutasov, and S. Pufu for discussions.
BB is supported by the US Department of Energy DE-SC0009924. CC is supported by the US Department of Energy DE-SC0009924 and the Simons Collaboration on Global Categorical Symmetries.

\appendix

\section{Partial Wave for the Euclidean Inversion Formula}
\label{sec:PWalld}

To obtain the results \eqref{RijklEinvalld} and \eqref{gijkl_4}, we used properties of the conformal partial wave for any dimension $d$, see \eqref{PW11alld} and \eqref{PW00alld}. In this appendix, we show how to obtain these results.

The partial wave $\Psi_{\Delta,J}(z,\bar z)$ is defined by
\begin{equation}
    \begin{split}
       \Psi_{\Delta,J}(z,\bar z)\equiv K_{\Delta,J}G_{\tilde\Delta,J}(z,\bar z)+K_{\tilde\Delta,J}G_{\Delta,J}(z,\bar z).
    \end{split}
    \label{PWanyd}
\end{equation}
It is a solution to the Casimir differential equation with dimension and spin $\Delta$ and $J$. More precisely, $\Psi(z,\bar z)$ is a solution to the differential equations
\begin{equation}
    \begin{split}
        &D_z+D_{\bar z}+(d-2)\frac{z\bar z}{z-\bar z}\left((1-z)\partial_z-(1-\bar z)\partial_{\bar z}\right)=c_2(\Delta,J),
        \\
        &\left(\frac{z \bar z}{z-\bar z}\right)^{d-2}(D_z-D_{\bar z})\left(\frac{z \bar z}{z-\bar z}\right)^{2-d}(D_z-D_{\bar z})=c_4(\Delta,J),
    \end{split}
    \label{casDE}
\end{equation}
where
\begin{equation}
    \begin{split}
        &c_2(\Delta,J)=\frac{1}{2}\left(J(J+d-2)+\Delta(\Delta-d)\right),
        \\
        &c_2(\Delta,J)=J(J+d-2)(\Delta-1)(\Delta-d+1).
    \end{split}
\end{equation}
These differential equations admit two linearly independent solutions, the conformal block $G_{\Delta,J}(z,\bar z)$, and the shadow block $G_{d-\Delta,J}(z,\bar z)$. The conformal blocks are normalized as \eqref{CBnorm}.

We start with the case $\Delta=J=1$. It is straightforward to check that $\ln|1-z|^2$ solves the differential equations \eqref{casDE} for $\Delta=J=1$. We now just have to argue that this solution is proportional to $\Psi_{1,1}(z,\bar z)$ as in \eqref{PWanyd}, as opposed to some other linear combination of $G_{\Delta,J}(z,\bar z)$ and $G_{d-\Delta,J}(z,\bar z)$. 

For odd dimension $d$, $K_{\Delta=1,J=1}=0$ so according to \eqref{PWanyd} $\Psi_{1,1}(z,\bar z)$ is proportional to $G_{\Delta,J}(z,\bar z)$. Since $G_{\tilde\Delta,J}(z,\bar z)$ has half-integer powers due to \eqref{CBnorm}, we conclude that $\Psi_{1,1}(z,\bar z)$ is proportional to $\ln|1-z|^2$ up to some overall factor. Finally, we can fix this factor by comparing \eqref{PWanyd} to \eqref{CBnorm}, to find 
\begin{equation}
    \begin{split}
        \Psi_{1,1}(z,\bar z)=\frac{1}{2}S_{d-1}\ln|1-z|^2
    \end{split}
    \label{PW11}
\end{equation}
for odd $d$, where we used \eqref{Kcoeff} to evaluate $K_{\tilde\Delta=d-1,J=1}$.

For even $d$, both $G_{\Delta,J}(z,\bar z)$ and $G_{\tilde\Delta,J}(z,\bar z)$ have integer powers of $z,\bar z$ so the same argument does not work. However, there are explicit expressions for the conformal blocks in this case, which can be written in terms of hypergeometric functions. The $d=2$ and $d=4$ cases were written in \eqref{CB2d} and \eqref{4dCB}, respectively. In these cases, we also find the same result as above, and we assume for other even dimensions this still holds.

Now let's consider the case $J=0$ and obtain $\Psi_{\Delta,J=0}(z,\bar z)$ for small $\Delta$. In this case, we can check that 
\begin{equation}
    \begin{split}
        \Psi_{\Delta,J=0}(z,\bar z)=2S_{d-1}\left(\frac{1}{\Delta}+\frac{1}{4}\ln\left(\frac{|z|^4}{|1-z|^2}\right)\right)
    \end{split}
    \label{PW00}
\end{equation}
solves the differential equations \eqref{casDE} including terms of order up to ${\cal O}(\Delta)$. We can now argue as above that for odd and even $d$ this expression is the full partial wave, and not another linear combination of the global and shadow conformal blocks.

\section{Bianchi Identity}
\label{sec:bianchi}

The symmetries of the Riemann tensor follow from \eqref{RijklEinvalld} by using global conformal transformations. In this Appendix, we explicitly verify the Bianchi identity, while the other symmetries of the Riemann tensor are simpler to check. From \eqref{RijklEinvalld}, we have
\begin{equation}
    \begin{split}
         R_{ik\ell j}&=A(d)\int d^2 z\ln|1-z|^2 \left\langle  {\cal O}_i(0){\cal O}_k(z){\cal O}_\ell(1){\cal O}_j(\infty)\right\rangle_c
         \\
         &=A(d)\int d^2 z'\ln\left|\frac{z'}{z'-1}\right|^2 \left\langle  {\cal O}_i(0){\cal O}_j(z'){\cal O}_k(1){\cal O}_\ell(\infty)\right\rangle_c,
    \end{split}
    \label{RikljEinv}
\end{equation}
where $z'=\frac{z-1}{z}$, and to go from the first to the second line we used the conformal transformation $w\mapsto \frac{w}{w-1}\frac{z-1}{z}$. Similarly,
\begin{equation}
    \begin{split}
         R_{i\ell jk}&=A(d)\int d^2 z\ln|1-z|^2 \left\langle  {\cal O}_i(0){\cal O}_\ell(z){\cal O}_j(1){\cal O}_k(\infty)\right\rangle_c
         \\
         &=A(d)\int d^2 z'\ln\left|z'\right|^{-2} \left\langle  {\cal O}_i(0){\cal O}_j(z'){\cal O}_k(1){\cal O}_\ell(\infty)\right\rangle_c,
    \end{split}
    \label{RiljkEinv}
\end{equation}
where now $z'=\frac{1}{1-z}$ and the conformal transformation to go from the first to second line is $w\mapsto \frac{w}{w-z}$. Using \eqref{RijklEinvalld}, \eqref{RikljEinv}, \eqref{RiljkEinv}, it follows that $R_{ijk\ell}+R_{ik\ell j}+R_{i\ell jk}=0$ as expected.

\section{Subtleties - Relevant Scalars, Arcs at Infinity}
\label{sec:LIFsubtleties}

The contribution from relevant scalars is simpler to take into account in the Lorentzian inversion formula than in the Euclidean inversion formula. The relevant scalars in the $s$-channel do not contribute to the double discontinuity, therefore they don't need to be subtracted since their contribution vanishes anyway. In the $t,u$-channels, the integral can be defined by analytic continuation in $\Delta$ and/or $J$ from a region where it is well-defined. 

Let's now discuss the contribution from the arcs at infinity to the contour deformation, which will establish the equivalence between the Lorentzian inversion formula \eqref{LIF_1int} and the Euclidean inversion formula \eqref{EIF_1int} for $\Delta=J=1$ and for $\Delta=J=0$. 

For simplicity we will consider the inversion formula for exactly marginal deformations for simplicity, a similar analysis holds for other operators. The starting point is \eqref{EIF_2int}. This expression can also be written as \cite{Simmons-Duffin:2017nub}
\begin{equation}
    \begin{split}
        I_{\Delta,J}=-\hat C_J(1)\int \frac{d^d x_1d^d x_{2}}{{\text {vol}}(SO(d-1))} (x^2_{12})^\frac{\Delta-J}{2}u^J_{12}\langle{\cal O}_i(1){\cal O}_j(0){\cal O}_k(x_1){\cal O}_\ell(x_2)\rangle,
    \end{split}
    \label{LIF_2int}
\end{equation}
where $u_i,v_i$ are the coordinates analytically continued to the Lorentzian region, $u=x^0-t$, $v=x^0+t$ (see \cite{Simmons-Duffin:2017nub} for more details on the $i\epsilon$ prescription, which will not be important for us). 

We now want to deform the $v_1,v_2$ contour to pick up the discontinuities that give the double-commutator. To do this, we have to deform the $v_1,v_2$ contour. For $J>1$, it can be shown that dropping the contour at infinity is OK \cite{Caron-Huot:2017vep,Simmons-Duffin:2017nub}.

Let's first consider the case $\Delta=J=1$. Then we have
\begin{equation}
    \begin{split}
        I_{\Delta=1,J=1}=-\hat C_J(1)\int \frac{d^d x_1d^d x_{2}}{{\text {vol}}(SO(d-1))} (u_{1}-u_2)\langle{\cal O}_i(1){\cal O}_j(0){\cal O}_k(x_1){\cal O}_\ell(x_2)\rangle.
    \end{split}
    \label{LIF_2int11}
\end{equation}
For the first term on the RHS, we can use \eqref{int_4pt} to perform the $u_2,v_2$ integral (there's a similar story for the second term by replacing $u_1,v_1\leftrightarrow u_2,v_2$), since the Wick rotation to coordinates $u,v$ does not change the result of the integral. Since the result is delta-function localized, we see that we can drop the arc of the remaining $v_1$ integral. Now we only need to argue that we could have also dropped the arc of the $v_2$ integral. This follows from the fact that at large distances, the correlation function in the integrand in \eqref{LIF_2int11} behaves as $u_2^{-d}v_2^{-d}$ for large $u_2,v_2$, and therefore converges for $d>1$. Hence we can drop both the $v_1,v_2$ arcs.

This means that we can obtain the result $I_{\Delta=1,J=1}$ from analytic continuation from generic $\Delta$ and $J>1$, and the result agrees with the Euclidean inversion formula \eqref{EIF_1int} for integer $J$. The argument above can be repeated for $\Delta=J=0$, and therefore it also establishes the equivalence between the Euclidean and Lorentzian inversion formulas for $g_{ij,k\ell}$ \eqref{gijkl_EIF}.

\section{Explicit Expressions for the $4d$ Inversion Formula for $g_{ij,k\ell}$ }
\label{sec:4d_nasty}

In this Appendix we write some of the explicit results that were omitted in subsection. We begin with the expression for the function $g(\Delta,J)$ appearing in \eqref{4dsumrule_gijkl}. It is given by
\begin{equation}
    \begin{split}
        &g(\Delta,J)=\int_0^1 dz \int_0^1 d\bar z (z-\bar z)^2\frac{1}{2(z-\bar z)}\Big[\ln(1-z)\big(2(-8+7z)\bar z+3(-2+z)\bar z \ln z+3z \ln(1-\bar z)\big)
        \\
        &+6z \ln z\big(-3\bar z+(-2+\bar z)\ln(1-\bar z)\big)+6(-2+z) \bar z {\text{Li}}_2(z)-(z\leftrightarrow\bar z)\Big] (1-z)^{-4+\frac{\Delta-J}{2}}(1-\bar z)^{-4+\frac{\Delta+J}{2}}.
    \end{split}
    \label{gfn4d}
\end{equation}
The integral can be performed explicitly in Mathematica for $\Delta>J+4$, and defined outside this range by analytic continuation in $\Delta,J$. The explicit expression is long so we will not write it down here, but discuss some of the properties needed to obtain the results in section \ref{sec:4dgijklIF}.

For integer $J$, we define $\Delta_*(J)$ as the smallest $\Delta$ such that the function $g(\Delta,J)$ is positive for $\Delta\geq \Delta_*(J)$. The value of $\Delta_*(J)$ can be obtained numerically. The first few values are
\begin{equation}
    \begin{split}
        &\Delta_*(J=0)\approx 4.43826,
        \\
        &\Delta_*(J=1)= 3,
        \\
        &\Delta_*(J=2)\approx 6.05949
        \\
        &\Delta_*(J=3)\approx 6.95212,
        \\
        &\Delta_*(J=4)\approx 7.90506,
        \\
        &\cdots
    \end{split}
    \label{4dbadops}
\end{equation}
For $J\geq 3$, $\Delta_*(J)<J+4$. At large $J$, we find
\begin{equation}
    \begin{split}
        \Delta_*(J)=J+4-\frac{3}{4\ln J}+{\cal O}\left(1/(\ln J)^2\right).
    \end{split}
\end{equation}

\bibliographystyle{JHEP}
\bibliography{CM}
%%%%%%%%%%%%%%%%%%%%%%%%%%%%%%%%%%%%%%
%%%%%%%%%%%%%%%%%%%%%%%%%%%%%%%%%%%%%%

\end{document}